\newif\ifsubmode
\submodefalse

\newif\ifprintfig
\printfigtrue

\ifsubmode
  \documentstyle[12pt,aasms4,epsf]{article}
  \received{}
  \accepted{}
  \journalid{}{}
  \articleid{}{}
\else
  \documentstyle[11pt,aaspp4,epsf]{article}
  \slugcomment{{\it Astronomical Journal, Jan 2004, in press}}
\fi

\lefthead{HST Observations of NGC 6240}
\righthead{Gerssen et al.}

\newcommand{\etal}{{et al.~}}

\newcommand{\lta}{\lesssim}
\newcommand{\gta}{\gtrsim}

\newcommand{\kms}{\>{\rm km}\,{\rm s}^{-1}}
\newcommand{\pc}{\>{\rm pc}}
\newcommand{\kpc}{\>{\rm kpc}}

\newcommand{\Msun}{\>{\rm M_{\odot}}}
\newcommand{\Lsun}{\>{\rm L_{\odot}}}

\newcommand{\rmh}{\rm h}
\newcommand{\rmm}{\rm m}
\newcommand{\rms}{\rm s}

\def\Halpha{{\,\rm H{\alpha}}}

\begin{document}

\title{Hubble Space Telescope Observations of NGC~6240: a Case Study of 
       an Ultra-Luminous Infrared Galaxy with Obscured Activity}

\author{Joris Gerssen, Roeland P.~van der Marel}
\affil{Space Telescope Science Institute, 3700 San Martin Drive, 
       Baltimore, MD 21218}

\author{David Axon} 
\affil{Department of Physics, Rochester Institute of Technology, 
84 Lomb Memorial Drive, Rochester, NY 14623}

\author{J. Christopher Mihos\altaffilmark{1}} 
\affil{Department of Astronomy, Case Western Reserve University, 10900 Euclid
Avenue, Cleveland, OH 44106}

\author{Lars Hernquist} 
\affil{Harvard University, Center for Astrophysics, 60 Garden Street,
MS-51, Cambridge MA 02138}

\author{Joshua E. Barnes} 
\affil{University of Hawaii, Institute for Astronomy, 2680 Woodlawn Drive,
Honolulu, HI 96822}

\altaffiltext{1}{Research Corporation Cottrell Scholar and NSF CAREER
Fellow}


\ifsubmode\else
\clearpage\fi


\ifsubmode\else
\baselineskip=14pt
\fi


\begin{abstract}
We present results from a Hubble Space Telescope (HST) study of the
morphology and kinematics of NGC 6240. This merging galaxy with a
double nucleus is one of the nearest and best-studied ultraluminous
infrared galaxies. HST resolves both nuclei into seperate components.
The distance between the northern and southern optical/near-infrared
components is greater than that observed in radio and X-ray studies,
arguing that even in K-band we may not be seeing all the way through
the dust to the true nuclei.

The ionized gas does not display rotation around either of the nuclei,
or equilibrium motion in general. There is a strong velocity
gradient between the nuclei, similar to what is seen in CO data. There
is no such gradient in our stellar kinematics. The velocity dispersion
of the gas is larger than expected for a cold disk. We also map and
model the emission-line velocity field at an off-nuclear position
where a steep velocity gradient was previously detected in
ground-based data. Overall, the data indicate that line-of-sight
projection effects, dust absorption, non-equilibrium merger dynamics,
and the possible influence of a wind may be playing an important role
in the observed kinematics.

Chandra observations of hard X-rays have shown that both of the nuclei
contain an Active Galactic Nucleus (AGN). The HST data show no clear
sign of the two AGNs: neither continuum nor narrow-band imaging shows
evidence for unresolved components in the nuclei, and there are no
increased emission line widths or rapid rotation near the nuclei. This
underscores the importance of X-ray data for identifying AGNs in
highly dust-enshrouded environments.
\end{abstract}





\section{Introduction}
\label{s:intro}

Ultraluminous Infrared Galaxies (ULIRGs) have infrared luminosities in
excess of $\sim 10^{12} \Lsun$. These galaxies were first discovered
by the IRAS satellite (e.g., Soifer, Neugebauer \& Houck 1987). They
are usually associated with mergers and interactions of galaxies
(e.g., Sanders \& Mirabel 1996) and their infrared emission is powered
a combination of a nuclear starburst and an embedded AGN (e.g. Genzel
\etal 1998). It is believed that the ULIRGs observed locally are 
related in an evolutionary sense to the onset of quasar activity
(Sanders \etal 1988), the formation of (elliptical) galaxies (Kormendy
\& Sanders 1992), the sub-millimeter sources observed with SCUBA
(Smail, Ivison, \& Blain 1997), and the many obscured AGNs detected
with Chandra (Barger \etal 2001).

NGC~6240 at a distance of 104 Mpc (assuming $H_0 = 70 \kms$
Mpc$^{-1}$, both here and throughout this paper) has $L_{\rm IR} =
10^{11.8} \Lsun$, just below the canonical ULIRG luminosity boundary
of $10^{12} \Lsun$. Nonetheless, it is generally regarded both as a
{\it bona fide} ULIRG, and as one of the nearest and best studied
examples of the class (Genzel \etal 1998). The large-scale optical
morphology of NGC 6240 is highly distorted and shows the tidal tails
indicative of an ongoing merger (Keel 1990; see also
Figure~\ref{f:DSS}). Ground-based imaging at both optical (Schulz
\etal 1993) and near-infrared wavelengths (Doyon \etal 1994) has
revealed a double nucleus with a separation of $\sim 1.8'' = 0.88
\kpc$. Radio continuum images also show the double nucleus (Carral, 
Turner, \& Ho 1990; Eales \etal 1990; Beswick \etal 2001), as do [Fe
II] images at $1.64\mu$ (van der Werf et al.~1993). 

The infrared emission from NGC 6240 comes from dust heating.
Mid-infrared line-ratio measurements with ISO show that the heating is
due partly to a nuclear starburst triggered by the interaction, and
partly due to an AGN continuum (Genzel \etal 1998). The optical
emission of NGC 6240 is characterized by a LINER-type spectrum and the
width of the H$\alpha$+[NII] emission lines increases strongly towards
the double nucleus (Keel 1990). The presence of an AGN component in
NGC 6240 was confirmed by the detection of hard X-rays with ASCA
(Iwasawa \& Comastri 1998) and BEPPO-SAX (Vignati \etal
1999). Interestingly, subsequent Chandra observations showed that
there is actually an AGN associated with each of the two nuclei of
NGC~6240 (Komossa \etal 2003). It is not necessarily surprising to
find two AGNs in a merger of two galaxies, given that most galaxies
are believed to harbor central black holes (e.g., Kormendy \& Gebhardt
2001) and given that mergers have been implicated as triggers for AGN
activity. However, NGC 6240 is the first galaxy for which the presence
of two supermassive black holes has been convincingly
demonstrated.\looseness=-2

The stellar velocity field in the region of the double nucleus has
been studied from the ground by Tecza \etal (2000). The velocity field
of molecular CO gas was studied by Tacconi \etal (1999). Neither of
these studies found a tell-tale signature of one of the black holes in
NGC 6240 in the form of unusually rapidly moving stars or
gas. However, this is not surprising given that the spatial resolution
of Hubble Space Telescope is generally required to detect the
gravitational influence of super-massive black holes in galaxies.

NGC 6240 is of considerable importance as a nearby ULIRG proto-type.
Its structure has therefore been studied in great detail at many
wavelengths. However, ground-based studies are generally limited to
resolutions near $\sim 1'' = 0.49 \kpc$. By contrast, HST can probe
down to scales of $\sim 0.05'' = 25 \pc$. We present here the results
of an HST study of the morphology and kinematics of NGC 6240 at
optical wavelengths, with the Second Wide Field and Planetary Camera
(WFPC2) and Space Telescope Imaging Spectrograph (STIS), respectively.
Previous HST studies of NGC~6240 were restricted to pre-COSTAR U-band
imaging with the Faint Object Camera (FOC) by Barbieri \etal (1993)
and Near-Infrared Camera and Multi-Object Spectrometer (NICMOS)
imaging by Scoville \etal (2000) and van der Werf (2001). The U-band
data showed multiple components in the nuclear region, probably due to
patchy dust obscuration. The NICMOS data confirmed the double nucleus.

ULIRGs, and galaxy mergers in general, tend to be more dusty than
normal galaxies. This can complicate investigations into their
structure. Near-IR or mid-IR observations are definitely best suited
to penetrate the dust. On the other hand, the observational
capablities at these wavelengths are not always comparable to what can
be achieved at optical wavelengths. For example, HST has an optical
spectrograph, but no near-IR spectrograph. Of course, optical
observations are best restricted to those galaxies for which one has a
chance to penetrate the dust. Ground-based optical observations of NGC
6240 have shown the same double nuclues structure seen in radio and
X-ray data. This suggests that optical observations may be seeing far
enough through the dust to be able to probe the regions where the
radio and X-ray emission are produced. With this in mind, we have used
HST to study NGC 6240 with the following goals:

\noindent (1) Improve our understanding of the nuclear structure and 
morphology of NGC 6240. Ground-based observations have found that the
separation between the two nuclei of NGC 6240 is somewhat wavelength
dependent (Schulz \etal 1993), which indicates that dust absorption is
not negligible. The spatial resolution of HST, combined with the
availability of data in various bands, allows us to study the
distribution and optical depth of the dust. Comparison of the observed
morphology to that at radio and X-ray wavelengths provides insight
into the extent to which the AGN components of NGC 6240 are, or are
not, obscured at optical and near-infrared wavelengths.

\noindent (2) Search for tell-tale signs of the AGN components in 
NGC 6240. For example, unresolved continuum emission might point to
optical synchrotron emission from one of the AGNs. Increases in
stellar or gaseous motions can pin-point the gravity of a black hole,
and if detected, might allow measurements of the masses of the black
holes in NGC 6240.

\noindent (3) Examine the extent to which the kinematical characteristics of 
NGC 6240 are consistent with relaxed quasi-equilibrium motions. This
is relevant for a variety of issues. For example, Bland-Hawthorn,
Wilson \& Tully (1991) suggested that the large-scale velocity field
can be decomposed into the contributions of two rotating
disks. Similarly, Tecza \etal (2000) suggested that the small-scale
velocity field near the double nucleus of the system can be modeled as
two rotating nuclear disks.  Conversely, Tacconi \etal (1999) fit the
nuclear CO velocity field with a single disk centered between the
nuclei. Lester \& Gaffney (1994) used the stellar velocity dispersion
to estimate the mass of NGC 6240, and Doyon \etal (1994) used it to
place NGC 6240 on the fundamental plane of elliptical galaxies.

In Section~\ref{s:imag} we present WFPC2 imaging of NGC 6240 in the
broad B, V, and I-bands, as well as narrow-band imaging of
$\Halpha$+[NII] emission. In Section~\ref{s:spec} we study the
small-scale velocity field of NGC 6240 using STIS Ca triplet
absorption-line spectroscopy and $\Halpha$+[NII] emission line
spectroscopy of the nuclear region. We also use $\Halpha$+[NII]
emission line spectroscopy to map and model the gas velocity field at
the off-nuclear position where a steep velocity gradient was
previously detected in ground-based data (Bland-Hawthorn \etal
1991). Conclusions are presented in Section~\ref{s:conc}. Preliminary
results of our project were discussed previously in Gerssen \etal
(2001). Also, some of our WFPC2 data have already been used in
archival studies by other groups. Our $\Halpha$+[NII] image was
discussed by Lira \etal (2002) and Komossa
\etal (2003) in the context of X-ray observations of NGC~6240. Our
broad-band images were used by Pasquali, de Grijs \& Gallagher (2003)
in a study of the star cluster population of NGC~6240.

\section{WFPC2 Imaging}
\label{s:imag}

\subsection{Observations and Data Reduction}
\label{ss:WFobsreduc}

Broad- and narrow-band images of NGC~6240 were obtained in March 1999
with the WFPC2 on board of HST in the context of program GO-6430 (PI:
van der Marel).  The WFPC2 is described in detail by, e.g., Biretta
\etal (2001).  The observations were obtained in a single `visit' 
of four spacecraft orbits.  The WFPC2 has four chips, each with $800
\times 800$ pixels. All exposures were obtained with the galaxy
centered on the PC chip to obtain the highest spatial resolution
(pixel size $0.046''$). Figure~\ref{f:DSS} shows the PC field of view
overlaid on an NGC 6240 image extracted from the Digitized Sky Survey.

The observing log is presented in Table~\ref{t:images}. Broad band
observations were obtained with the F450W, F547M and F814W filters
(roughly Johnson $B$, $V$ and $I$). Narrow band observations were
obtained using the F673N and the F658N filters. At the redshift of
NGC~6240 ($z=0.02427$) the $\Halpha$+[NII] lines are shifted into the
wavelength range of the F673N (`on-band') filter. NGC~6240 is
emission-line free in the F658N (`off-band') filter.

The images were calibrated by the HST calibration `pipeline'
maintained by the Space Telescope Science Institute (STScI). The
standard reduction steps include bias subtraction, dark current
subtraction and flat-fielding, as described in detail by Holtzman
\etal (1995a). Two or three different exposures were taken with each 
filter, each with the target shifted by a small integer number of
pixels. The different exposures for each filter were aligned, and then
combined with simultaneous removal of cosmic rays and chip
defects. This yields one final image per filter. To obtain an image of
the $\Halpha$+[NII] emission we aligned the off-band image with the
on-band image, scaled it to fit the on-band image at radii where the
ionized gas flux is negligible, and subtracted it from the on-band
image.

The count-rates in the broad-band F450W, F547M and F814W images were
calibrated to magnitudes in the Johnson $B$, $V$ and~$I$ bands,
respectively, as described in Holtzman \etal (1995b). The count-rates
in the F673N image were calibrated to erg cm$^{-2}$ s$^{-1}$ using
calculations with the SYNPHOT package in IRAF. 

\subsection{Global Morphology}
\label{ss:golbmorph}

Figure~\ref{f:PCfield} shows a composite color image of the three
broad band images for the full PC field of view, with the
$\Halpha$+[NII] image overlaid in yellow. The broad-band data show a
complex morphology, consistent with the assumed merger nature of NGC
6240. A wide dust lane is visible against the main body of the galaxy,
which was studied previously by Keel (1990) using unsharp masking. The
WFPC2 image shows additional filamentary dust structures on smaller
scales. These extend into the very central region and (partly) cover
the nuclear region. 

Heckman, Armus \& Miley (1987, 1990) found that a morphologically
spectacular emission-line nebula is associated with NGC 6240. They
interpreted this as evidence for a 'superwind'. The kinetic energy
provided by supernovae and winds from massive stars in a central
starburst drives a large-scale outflow that can shock-heat and
accelerate ambient interstellar and circumgalactic gas.
Figure~\ref{f:PCfield} reveals the intricate nature of the superwind
at high spatial resolution. The emission-line filaments (yellow)
appear to trace out a bipolar outflow pattern, with an axis aligned
roughly E-W, perpendicular to the dust lane and the line connecting
the two nuclei. This morphology is similar to that seen in other
starburst superwind galaxies. The curved filament at the top of the
image may be due to expelled gaseous material that is now falling back
towards the central parts of the galaxy. Extended X-ray (0.5--1.5 keV)
images (Lira \etal 2002; Komossa \etal 2003) have a morphology similar
to the $\Halpha$+[NII] image, as would be expected if the origin of
the filaments is a superwind. In particular, Komossa \etal (2003)
report several X-ray loops and knots that have a corresponding bright
structure in our WFPC2 gas emission line image. The connection between
the X-ray observations and our $\Halpha$+[NII] image was already
discussed in detail by Lira \etal (2002) and Komossa \etal (2003), so
we will not further discuss it here.\looseness=-2

\subsection{Nuclear Structure}
\label{ss:nucstruc}

Figure~\ref{f:PCblowup} shows the central region of NGC 6240 at high
contrast. Figure~\ref{f:PCblowup}a shows a true-color image of the
broad-band WFPC2 data, and Figure~\ref{f:PCblowup}b shows a gray scale
image of the $\Halpha$+[NII] data. Previous studies of NGC 6240 have
revealed that it has a double nucleus. The new HST images show that
the central structure is in fact more complicated. Instead of two,
there are three distinct nuclear components, both in broad-band and in
$\Halpha$+[NII] gas emission. We will refer to these components as N1,
N2 and N3, as indicated in Figure~\ref{f:PCblowup}a. N1 corresponds to
the northern nucleus that had been identified from ground-based data,
while N2 and N3 jointly correspond to the southern nucleus identified
from ground-based data.  The distance between N2 and N3 is $0.45''$,
which explains why these components were not spatially resolved in
arcsec-resolution ground-based data.

We determined the magnitudes of the nuclear components in each of the
three available broad-band filters. For each nucleus we used a
circular aperture with a radius of 4 pixels ($0.182''$). The results
are listed in Table~\ref{t:nucmags}. The magnitudes themselves are not
of particular interest, since they depend strongly on the adopted
aperture size. The implied colors are of more direct physical
interest.  N2 is the reddest of the three nuclei, as is evident also
from Figure~\ref{f:PCblowup}a. The centroid of the combined light of
N2 and N3 varies as a function of wavelength, because N2 and N3 do not
have the same color. In the $B$-band the centroid is close to N3,
while in the $I$-band it is between N2 and N3. This explains the
ground-based data of Schultz \etal (1993), which show that the
separation between the northern and the southern nucleus of NGC 6240
decreases with increasing wavelength.

We analyzed a NICMOS $K$-band image from the HST archive (obtained
February 1998, program GO-7219, PI: Scoville) in order to further
improve our understanding of the nuclear structure. A gray-scale image
of the $K$-band data is shown in Figure~\ref{f:PCblowup}c, at the same
scale and orientation as Figures~\ref{f:PCblowup}a,b. This K-band
image was discussed previously, in combination with NICMOS J and
H-band data, by Scoville \etal (2000). The $K$-band morphology shows
only two main components, unlike the optical images. The brightest
pixel in the elongated southern component of the $K$-band image
coincides with N2, confirming that N2 is indeed the true center of the
southern component.  The N3 component cannot be identified on the
$K$-band image, although its location coincides with the extension of
the southern component.  N3 is therefore most likely either an HII
region or an artifact resulting from the highly patchy dust structure
in the nucleus of NGC~6240, rather than a physically distinct nuclear
component. Note that the FWHM of the point spread function (PSF) is
much larger for the K-band data (0.19 arcsec) than for, e.g., the
V-band data (0.06 arcsec; these PSF FWHM values were calculated with
the TinyTim software of Krist \& Hook 2001). This is (at least part
of) the reason that the K-band morphology looks smoother than the
optical morphology. However, the K-band FWHM is much less than the
distance between N2 and N3 ($0.45''$). The fact that N3 cannot be
identified as a distinct peak in the K-band is therefore not an
artifact of limited resolution. To facilitate a comparison with the
optical results we determined the $K$ band magnitudes in the same
three apertures used for the WFPC2 photometry. The results are listed
in Table~\ref{t:nucmags}.

It is interesting to note that the northern N1 component in the
$K$-band image is much more elongated than in the $I$-band. It appears
that the southern extension to the N1 component that is visible in the
K-band data (indicated as N1$'$ in Figure~\ref{f:PCblowup}c) could in
fact be a separate component. This suggestion is strengthened by a
Keck adaptive optics K-band image of the central region of NGC 6240,
obtained by C.~Max and shown in Schneider (2002). That image has
slightly higher spatial resolution than the NICMOS image. If indeed
N1$'$ is a separate component it must have high extinction, given that
it does not show up in the I-band data.

The amount of dust absorption towards the nuclear components can be
estimated if one assumes that the dust is in front of the nuclei, and
if one has estimates of the intrinsic colors.  The largest color
baseline line of the data is provided by the $B-K$ colors. The $A_V$
can be calculated from the color excess $E_{B-K} = (B-K)_{\rm obs} -
(B-K)_{\rm intrinsic}$. The extinction law of Rieke \& Lebofsky (1985)
yields $A_V = 0.825 E_{B-K}$. The average $B-K$ colors for late-type
spiral, early-type spiral and elliptical galaxies are 3.18, 3.75 (de
Jong 1996) and 4.15 (Persson, Frogel \& Aaronson 1979; Carollo \etal
1997) respectively. If any of these are taken as estimates of the
intrinsic colors of the NGC 6240 nuclei, then this yields three
estimates of $A_V$. Table~\ref{t:nucmags} lists the average of the
estimates thus obtained. The error bars indicate the range spanned by
the calculated values. These derived $A_V$ estimates are lower limits
because young stellar populations can be much bluer than the global
colors of normal galaxies. Knowledge of the stellar population age
near the center of NGC 6240 is available from other work, but
unfortunately this does not help much in constraining the intrinsic
$B-K$ color. Tecza
\etal (2000) used the CO 2-0 absorption bandhead equivalent width to
estimate the burst age at 20 Myr. At this age, the `Starburst99'
software of Leitherer \etal (1999; available at
http:/$\!$/www.stsci.edu/science/starburst99/) predicts for a
single-burst population of solar metallicity and Salpeter IMF that
$B-K \approx 1.6$. On the other hand, for a 10 Myr old population the
predicted color is $B-K \approx 3.3$ and for a 5 Myr old population it
is $B-K \approx 0.3$. It is unclear whether the population of NGC 6240
is understood well enough to confidently choose among these ages and
colors. The bluest that the population could reasonably be is $B-K
\approx 0.0$, in which case the $A_V$ values in Table~\ref{t:nucmags}
would need to be increased by $\sim 3$.  Because of the large
uncertainties in our estimates of $A_V$ that already result from our
ignorance of the intrinsic $B-K$ colors, we have not attempted to
correct for the differences in the PSF between the $B$- and $K$-band
data. These corrections are not expected to be large because we have
chosen to adopt an aperture size for the photometry that exceeds the
PSF FWHM in both bands. The values $V_{\rm corr}$ in
Table~\ref{t:nucmags} list the V-band magnitudes of the nuclei
corrected for our estimated dust absorption, $V_{\rm corr} = V - A_V$.
The results indicate that N2 is the intrinsically brightest nuclear
component, as already suggested by the K-band image.

Tecza \etal (2000) recently estimated the dust absorption towards NGC
6240 from near-IR spectra in the $K$-band. Assuming, as do we, that
the dust resides in a foreground screen they find $A_V = 1.6$ for the
northern nucleus. This is somewhat lower than our value $A_V (N1) =
2.35$. For the southern nucleus they find $A_V = 5.8$, which exceeds
the values $A_V (N2) = 4.75$ and $A_V (N3) = 3.65$ inferred here. This
discrepancy is presumably due at least in part to the different
spatial resolution of their data ($0.8''$--$1.0''$). It is clear from
Figure~\ref{f:PCblowup} that the dust absorption in NGC 6240 is quite
filamentary and spatially variable, so $A_V$ estimates for different
spatial regions need not necessarily agree. Tecza \etal point out that
$A_V$ would be higher by a factor $\sim 2.5$ if the dust were
well-mixed, instead of in the foreground. If this is indeed the case,
this would bring our estimates more in line with other estimates in
the literature. We also caution that our measurements only probe to
the depth at which the nuclei become optically thick in K. If we are
not seeing all the way into the nuclei, measurements at longer
wavelengths will result in higher inferred values of $A_V$. For
example, Rieke \etal (1985) adopted $A_V \approx 15$ based on the
continuum levels in $K$ and $L$ spectrophotometry. Genzel \etal (1998)
listed extinction values for a number of ULIRGs based on line ratios
obtained from ISO SWS data, yielding typical $A_V$ values ranging from
5 to 45, and a value specifically for NGC~6240 of $A_V \ge 5$.
However, it should be kept in mind that all these estimates refer to
considerably larger spatial regions in NGC 6240 than the estimates
derived here from the HST data. On smaller scales, Vignati
\etal (1999) estimated from BeppoSAX X-ray observations of NGC~6240
that the HI column density is about $10^{21}$ to $10^{22}$ atoms
cm$^{-2}$, from which they infer $A_V \sim 10$.

The morphology of the H$\alpha$+[NII] emission from the nuclear region
(Figure~\ref{f:PCblowup}b) is similar to what is seen in the optical
broad-band images. The three components N1, N2 and N3 can be clearly
distinguished. The last column in Table~\ref{t:nucmags} reports the
$\Halpha$+[NII] emission line fluxes determined in the same apertures
as used for the broad band images. The N1 component contains the most
flux.

For all images we estimated the sizes of the nuclear components using
the tools in the IRAF task {\tt imexamine}. In each waveband we
determined the FWHM of an azimuthally averaged radial profile centered
on either N1, N2 or N3. In all instances the derived FWHM was larger
than the FHWM of the PSF in that particular waveband. The nuclei are
therefore resolved. This is not surprising, given that the extended
stellar distribution dominates the broad-band light and that star
formation over extended regions contributes to the emission line flux.
On the other hand, it would have been interesting if any of the nuclei
had contained an unresolved component. This could have pinpointed the
site of an AGN. An emission line point source could be an unresolved
narrow line region or broad line region. A broad-band point source
could be emission from the AGN or its surrounding torus, although such
sources are generally found only in radio-loud galaxies (Chiaberge,
Capetti \& Celotti, 1999; Verdoes Kleijn \etal 2002a). As it stands,
the ability to detect the presence of an AGN in optical imaging
remains extremely difficult in these very dusty ULIRGs.\looseness=-2

The HST data yield the following J2000 coordinates for the nuclei:
$(16\rmh \, 52\rmm \, 58.914\rms, \> 2^{\circ} \, 24' \, 3.64'')$ for
N1; $(16\rmh \, 52\rmm \, 58.900\rms, \> 2^{\circ} \, 24' \, 3.50'')$
for N1$'$; $(16\rmh \, 52\rmm \, 58.859\rms, \> 2^{\circ} \, 24' \,
1.95'')$ for N2; and $(16\rmh \, 52\rmm \, 58.872\rms, \> 2^{\circ} \,
24' \, 1.60'')$ for N3. For N1, N2 and N3 these coordinates were
measured from the pipeline-reduced I-band image. For N1$'$, which is
not seen in the I-band image, the coordinates were measured from the
pipeline-reduced K-band image, after aligning it to the I-band
image. In an absolute sense, the listed coordinates are accurate only
at the level of $\sim 1''$, which is the absolute pointing accuracy of
HST. However, the relative positions of the components are very
accurate, $\lta 0.04''$. The morphology of the nuclear structure does
depend on wavelength, so there may be additional systematic errors due
to spatially variable dust absorption.

NGC 6240 shows two nuclear components in radio continuum
emission. Carral, Turner \& Ho (1990) and Eales \etal (1990) list
coordinates of the components that are in mutual agreement to within
$0.1''$. The average of their results is $(16\rmh \, 52\rmm \,
58.923\rms, \> 2^{\circ} \, 24' \, 4.66'')$ for the Northern component
and $(16\rmh \, 52\rmm \, 58.886\rms, \> 2^{\circ} \, 24' \, 3.27'')$
for the Southern component. These coordinates differ at the $\sim 1''$
level from the coordinates derived from the HST data, as expected on
the basis of the absolute accuracy of the latter. The radio components
are separated from each other by $1.48''$ along position angle ${\rm
PA} = 21.3^{\circ}$ (measured from North over East). Recent
observations with MERLIN at 5 GHz (Beswick \etal 2001) are consistent
with this result. The angular resolution of these data is about 0.1
arcsec and the separation between the two brightest components seen
in the MERLIN data is $1.52''$. As discussed in Section~\ref{s:intro},
Chandra data have recently shown that NGC~6240 also has a double
nucleus in hard X-rays.  Komossa \etal (2003) quote the distance
between the X-ray nuclei as $1.4''$, consistent with the distance
between the radio nuclei to within the observational errors. The
southern nucleus is the brightest one, both at radio and at X-ray
wavelengths. The K-band image in Figure~\ref{f:PCblowup}c shows that
the same is true at near-IR wavelengths.

It it puzzling that the separation determined from the radio data and
the X-ray emission does not agree with the separation of the optical
HST components. N1 and N2 are separated by $1.86''$ along ${\rm PA} =
25.0^{\circ}$, while N1 and N3 are separated by $2.12''$ along ${\rm
PA} = 16.0^{\circ}$. This implies that at least one of the radio/X-ray
components does not have an optical counterpart, and conversely, that
at most one of the optical components has a radio/X-ray counterpart.
The K-band image shows a hint of a component N1$'$ to the south of N1
(Figure~\ref{f:PCblowup}c) that has high extinction. It could be that
that component marks the position of the northern radio/X-ray nucleus,
but even that does not fully resolve the discrepancy. N1$'$ and N2 are
separated along ${\rm PA} = 21.6^{\circ}$ by $1.67''$, which is still
larger than the nuclear separation seen in the radio and X-ray
data. Especially for the X-ray nuclei, there is strong (spectral)
evidence that they represent two AGNs (Komossa \etal 2003). Since
super-massive black holes are generally found in the regions of
highest stellar density, this suggests strongly that even in the
K-band we may not be seeing all the way through the dust.\looseness=-2

\section{STIS Spectroscopy}
\label{s:spec}

\subsection{Observations and Data Reduction}
\label{ss:specobs}

Long-slit spectra of NGC 6240 were obtained with STIS on HST between
May 1999 and June 2000 in the context of HST program GO-8261 (PI:
van~der~Marel). The STIS instrument is described in detail in Proffitt
\etal (2002). A total of ten orbits were available to study the
kinematics of the galaxy. Half of the time was used to study the
nuclear region. The other half was used to study a location about 12
arcsec east of the nuclear region where a steep gradient in the
velocity field of the ionized gas was previously detected in
ground-based data. For the nuclear region we obtained both
emission-line spectra and absorption line spectra. These spectra
totaled 2 and 3 orbits, respectively, and were all obtained in a
single visit. For the off-nuclear region we only obtained
emission-line spectra. These were obtained in two separate visits of 2
and 3 orbits, respectively. All spectra were obtained using the G750M
grating, which has a dispersion of $0.56${\AA} per pixel. The spatial
scale along the slit is $0.05''$ per pixel. The emission line spectra
cover the spectral range 6484--7054{\AA}, centered approximately on
the redshifted wavelength of $\Halpha$+[NII]. The associated velocity
scale is $24.8 \kms$ per pixel. The absorption line spectra cover the
spectral range 8539--9111{\AA}, centered approximately on the
redshifted wavelength of the redshifted Ca II triplet. The associated
velocity scale is $19.0 \kms$ per pixel.

To position the slits we started each visit with a target acquisition
on an uncatalogued star of magnitude $V=20.8$ located $18''$ from the
nuclear region of NGC 6240. This star has J2000.0 coordinates $(16\rmh
\, 52\rmm \, 59.847\rms, \> 2^{\circ} \, 24' \, 12.60'')$, to within
the $\sim 1''$ accuracy of the HST Guide Star Coordinate System. The
standard {\tt ACQ} acquisition procedure was used with a 100 sec
exposure time. This yields a nominal positional accuracy of $\lta
0.02''$ (Proffitt \etal 2002). The telescope was subsequently slewed
to the position of interest. The commanded slews were measured from
the WFPC2 images to $\lta 0.01''$ accuracy. The accuracy with which
small slews are executed by HST makes a negligible contribution to the
overall positional error budget.

The spectra were calibrated by the HST calibration `pipeline'
maintained by STScI. The standard reduction steps include bias
subtraction, dark current subtraction, flat-fielding,
wavelength-calibration to units of {\AA}, flux calibration to units of
erg cm$^{-2}$ s$^{-1}$ {\AA}$^{-1}$, and two-dimensional rectification
to linear spatial and wavelength scales. At each slit position two or
more exposures were obtained. These were dithered by an integer number
of pixels in the direction parallel to the slit, to allow removal of
CCD defects during data reduction. Different exposures for the same
slit position were aligned, and then combined with simultaneous
removal of cosmic rays and chip defects. The wavelength calibration
was performed using arc-lamp spectra obtained in the same visits as
the science spectra. This yields wavelength scales accurate to
0.1--0.2 spectral pixels, i.e., $\lta 5 \kms$.

The emission-line spectra were analyzed by simultaneous fitting of
Gaussian profiles to the emission lines using the software described
in van der Marel \& van den Bosch (1998). The wavelength range covered
by the spectra includes five emission lines: $\Halpha$ at a vacuum
rest-wavelength of $6564.6${\AA}, [NII] lines at $6549.8${\AA} and
$6585.3${\AA}, and [SII] lines at $6718.3${\AA} and $6732.7${\AA}. The
$S/N$ of the [SII] lines was generally low and in the following we
discuss only the $\Halpha$+[NII] lines. The Gaussian fits yield the
flux, mean radial velocity, and Gaussian velocity dispersion ($\sigma
= {\rm FWHM}/2.355$) of the lines.

The absorption-line spectra were rebinned onto a logarithmic scale and
were then analyzed using software that performs a fit in pixel space
(van der Marel 1994). The galaxy spectrum was modeled as the
convolution of a template spectrum and a Gaussian line-of-sight
velocity profile. The signal-to-noise ratio of the absorption line
spectra was insufficient to measure deviations of the velocity profile
shapes from a Gaussian, so no attempt was made to measure higher-order
Gauss-Hermite moments. The best-fit parameters of the Gaussian
velocity profile were determined by minimizing the residuals of the
fit in a $\chi^2$ sense. This yields the mean stellar velocity, the
velocity dispersion, and a line-strength parameter (a measure of the
equivalent width of the absorption lines in the galaxy spectrum with
respect to those in the template spectrum). The analysis uses all
absorption lines in the spectrum, but most of the signal is contained
in the three Ca II absorption lines at vacuum rest-wavelengths of
$8500.4${\AA}, $8544.4${\AA} and $8664.5${\AA}.  Template spectra were
not observed as part of our project, so we used a STIS spectrum of the
star HR~7615 (type K0III) from the HST Data Archive (project GO-7566,
PI: Green).

\subsection{Kinematics of the Nuclear Region}
\label{ss:kinnuc}

\subsubsection{Gas Kinematics}
\label{ss:gasnuc}

The gas kinematics of the nuclear region was studied by obtaining
spectra for seven different slit positions. The slits were aligned
parallel to line that connects the nuclei N1 and N2, using a pattern
that covers a contiguous area with a width of $1.1''$. The
lay-out of the slit positions is shown schematically in
Figure~\ref{f:slits}. The central three positions in the pattern were
observed with a slit of $0.1''$ width ({\tt 52x0.1}). Two exposures of
250 sec were obtained at each slit position, and no on-chip binning
was applied. A wider slit of $0.2''$ width ({\tt 52x0.2}) was used for
the outer four positions. Two exposures of 300 sec were obtained at
each slit position, and two-pixel on-chip binning was applied in both
the spatial and spectral directions (yielding effective pixels of
$0.10''$ by $1.12${\AA} $= 49.6 \kms$). This strategy was adopted to
increase the signal-to-noise ratio ($S/N$) of the spectra in areas
where the emission line flux is low. The spectral resolution for an
extended source observed with the $0.1''$-wide slit is $\sim 21 \kms$
(Gaussian dispersion), and for the $0.2''$-wide slit it is $\sim 42
\kms$.\looseness=-2

The spectra were analyzed by fitting Gaussians to the emission lines
as described above. In principle, each emission line is described by
three free parameters. However, the $S/N$ of the spectra is relatively
low and the lines are partially blended. We therefore found that more
robust results could be obtained by assuming that all emission lines
have the same mean velocity and velocity dispersion. This need not be
completely true in reality, but is a reasonable assumption in the
present context. We also fixed the ratio of the fluxes in the two
[NII] lines to the ratio of their transition probabilities (i.e., 3;
see references in Mendoza 1983). The ratio of the $\Halpha$ and [NII]
fluxes was kept as a free parameter. The best-fitting Gaussian
emission-line parameters were determined as a function of position
along the slit. For the observations with the $0.1''$ slit we applied
two-pixel binning along the slit to increase the $S/N$. The final
sampling along the slit was therefore $0.1''$ for both the narrow-slit
and the wide-slit observations. 

The emission-line fit results for the different slit positions were
combined and interpolated to yield two-dimensional maps of the flux,
radial velocity and velocity dispersion of the nuclear region of NGC
6240. The results are shown as two-dimensional greyscale and contour
maps in Figure~\ref{f:kinematics}. To get a different sense of these
data, Figure~\ref{f:gaskin} shows the projection along the $Y$-axis of
Figure~\ref{f:kinematics}. To obtain the latter results we summed the
fluxes obtained for the seven different slit positions and analyzed
the results with the same Gaussian-fitting method as described
above. Hence, Figure~\ref{f:gaskin} shows the kinematics for a
hypothetical $1.1''$-wide slit through N1 and N2, as function of
position along the slit. Here and throughout this paper we subtracted
the systemic velocity $v_{\rm sys} = 7287 \pm 8 \kms$ of NGC 6240
(Falco \etal 1999) from the mean velocity measurements.

The flux distribution inferred from the spectra
(Figure~\ref{f:kinematics}a) corresponds well with the $\Halpha$+[NII]
narrow band image (Figure~\ref{f:PCblowup}b; which is rotated with
respect to Figure~\ref{f:kinematics}a). The two nuclei are clearly
visible. The northern nucleus N1 (left in Figure~\ref{f:kinematics})
has the higher emission-line flux. The telescope was commanded to
place N2 at the origin of the coordinate system of
Figure~\ref{f:kinematics}. It is therefore clear that the offset-star
target acquisition and telescope slew procedures described in
Section~\ref{ss:specobs} were accurate to better than $\sim 0.03''$,
as expected.

The ionized gas velocity field does not show clear signs of rotation
around either of the two nuclei, or equilibrium motion in
general. Both N1 and N2 appear to be located in regions of lower mean
velocity than their surroundings.  The velocity of N2 is $\sim 250
\kms$ lower than that of N1. The highest mean velocity is seen between
N1 and N2, and at this point the mean velocity is $600 \kms$ larger
than the mean velocity of N2. A similar velocity gradient between the
nuclei was reported from observations of CO gas by Tacconi \etal
(1999), and was seen also in HI absorption by Beswick \etal
(2001). While Tacconi \etal found the CO velocity field to be highly
complex, they were able to fit it with a model of a rotating disk
between the two nuclei. The mean velocity profile in
Figure~\ref{f:gaskin}b is not inconsistent with this picture. The
profile is approximately antisymmetric around a position between N1
and N2, suggestive of rotation. However, the two-dimensional map in
Figure~\ref{f:kinematics}b is quite complex and it is unlikely that it
can be fit with any simple equilibrium model. For the case of the CO
velocity field, the suggestion of a rotating disk between the nuclei
received some support from the fact that the observed CO flux actually
peaks between the two nuclei. On the other hand, the ionized gas
emission that we observe clearly does not peak between the two nuclei
(see Figures~\ref{f:kinematics}a and~\ref{f:gaskin}a). This makes it
hard to explain why the ionized gas and the molecular gas have similar
kinematics. More generally, gas disks have often been argued to exist
around the nuclei of ULIRGs (Sakamoto \etal 1999), but only rarely
between them. Current N-body simulations do not yet have sufficient
resolution to follow the gaseous and stellar kinematics inside the
central kpc of mergers in much detail. However, they do indicate
that the dynamics can be complex and out of equilibrium, and that
projection effects from multiple kinematic components can lead to very
disorganized kinematics (Mihos 2000). Indeed, such complex velocity
fields from overlapping kinematic components have been observed in
other ULIRGs (Mihos \& Bothun 1998) and can only be disentangled
through two-dimensional mapping of the velocity fields with high
spectral resolution. An interpretation of the gas velocity curve of
NGC 6240 (Figure~\ref{f:gaskin}b) as simply being due to equilibrium
rotation in a organized disk may therefore be inappropriate. In view
of this, we have refrained from constructing models for the velocity
curve to attempt a determination of the central mass distribution of
the galaxy.

The velocity dispersion of the ionized gas is $\sim 300 \kms$ at both
N1 and N2. The velocity dispersion between the nuclei varies, dropping
to values near $\sim 200 \kms$. Overall, the velocity dispersion shows
a complex morphology and is larger than what would be expected for a
cold disk --- in some cases rising to $> 400\kms$. This indicates that
there is a strong non-equilibrium or turbulent component to the gas
kinematics, not entirely surprising in view of the merging nature of
this system. Given the superwind morphology of the ionized gas, the
features observed in the velocity maps may arise from outflow
kinematics. In a bipolar outflow, one would expect split emission
lines (e.g., Heckman \etal 1990); however, at the low $S/N$ of our
STIS data we would be unable to detect this splitting, particularly if
the outflow cone is filled. In this case, we would only measure an
increased velocity dispersion in the gas. Interestingly we also
measure the gas to be alternately red- and blue-shifted by about 300
$\kms$ along the slit when compared to the stellar velocities (see
Section~\ref{ss:starnuc} and Figure~\ref{f:stellarkin} below),
possibly consistent with a bipolar outflow scenario. However, velocity
mapping at higher spectral resolution with better two-dimensional
coverage than available here is needed to fully address the kinematic
state of the ionized gas in NGC 6240.

At the site of the nuclei themselves, we see no dramatic increase in
emission line width. Such an increase is often seen in gas kinematical
studies of AGNs with HST, even in galaxies that don't have a
broad-line region in the traditional sense (e.g., van den Bosch
\& van der Marel 1998; Verdoes Kleijn \etal 2000, 2002b). Had this
been seen in NGC 6240 as well, it would have provided a clear sign of
the AGN(s) responsible for the hard X-rays observed with
Chandra. However, even at HST resolution the ionized gas kinematics do
not a show a clear black hole signature.

\subsubsection{Stellar Kinematics}
\label{ss:starnuc}

Analysis of stellar kinematics requires higher S/N than analysis of
gas kinematics, given that absorption lines generally have smaller
equivalent widths than emission lines. With the available observing
time it was therefore not possible to study the stellar kinematics for
a two-dimensional region, as we did for the gas kinematics. Instead,
the $0.2''$-wide slit was used for only a single slit position. The
corresponding spectral resolution for an extended source is $\sim 32
\kms$ (Gaussian dispersion). The slit was placed at the center of the
pattern shown in Figure~\ref{f:slits}, thus covering both of the
nuclei N1 and N2. Six 1380 sec exposures were obtained, which were
co-added during analysis. To reduce the effects of read-out noise,
two-pixel on-chip binning was applied in both the spatial and spectral
directions (yielding effective pixels of $0.10''$ by $1.12${\AA} $=
38.0 \kms$).

The top panel of Figure~\ref{f:stellarkin} shows the continuum
intensity as function of position along the slit. The nuclei N1 and N2
are clearly visible. Away from the nuclei, the $S/N$ in the individual
rows of the two-dimensional spectrum was rather low. In these regions
we co-added spectra along the slit to increase $S/N$.  This procedure
yielded enough $S/N$ to determine the profile along the slit of the
mean stellar velocity, but insufficient $S/N$ to determine the profile
of the stellar velocity dispersion. However, it was possible to
determine a grand-total average velocity dispersion for the central
region. Co-addition of all data along the slit with the $x$-coordinate
of Figure~\ref{f:stellarkin} between $-3''$ and $1''$ yields $\sigma =
200 \pm 40 \kms$. These results are in adequate agreement with
ground-based measurements of Tecza \etal (2000), who analyzed
ground-based observations of the CO bandhead. They found $\sigma = 174
\pm 54 \kms$ for N1 and $\sigma = 236 \pm 24 \kms$ for N2. Tecza \etal
(2000) found a hint in their data that the stellar velocity dispersion
may be somewhat higher between the two nuclei, $\sigma = 276 \pm 51
\kms$, but the statistical significance of this result is low.

Both our results and those of Tecza \etal (2000) indicate that the
stellar velocity dispersion in the nuclear region is considerably less
than the values of $\sim 350 \kms$ that were obtained from earlier
ground-based observations (Lester \& Gaffney 1994; Doyon \etal
1994). It is possible that these larger values were due to the
combined effects of the poorer spatial resolution of these
observations and the nuclear gradients in mean velocity reported by
Tecza \etal (2000). Lester \& Gaffney (1994) used the perceived high
value of $\sigma$ to argue that NGC 6240 is quite massive, while Doyon
\etal (1994) used is to argue that NGC 6240 falls roughly on the
fundamental plane of elliptical galaxies (supporting the notion that
it might be an elliptical galaxy in formation). These arguments are
weakened by the lower $\sigma$ values found here with HST and
previously by Tecza \etal (2000). More importantly, $\sigma$ may be
influenced considerably by the non-equilibrium kinematics of the
merger, and it is risky to use it as an equilibrium measure of mass
and fundamental-plane position.\looseness=-2

To determine the profile of the mean stellar velocity along the slit
we performed Gaussian velocity profile fits for which the velocity
dispersion was kept fixed to the aforementioned value. Although this
ignores possible variations in $\sigma$ along the slit, this does not
affect the inferred mean velocities. This is because, to lowest order,
the mean and dispersion of a Gaussian are statistically uncorrelated
quantities (van der Marel \& Franx 1993). The solid symbols in the
bottom panel of Figure~\ref{f:stellarkin} shows the inferred mean
stellar velocities as function of position along the slit. For
comparison, the ionized gas velocities from Figure~\ref{f:gaskin} are
overplotted as open symbols. It is clear that the stellar velocities
differ considerably from the gas velocities. The stars do not show the
large velocity gradient between the nuclei that is seen for both the
ionized gas and for the cold CO gas (Tacconi
\etal 1999). The mean stellar velocities of N1 and N2 are similar. 
By contrast, the mean emission line velocity of N1 is $\sim 250 \kms$
lower than for N2. The fact that the stars and gas associated with the
nuclei of NGC 6240 do not have the same mean velocities indicates that
line-of-sight projection effects, dust absorption, and non-equilibrium
merger dynamics may be playing an important role in the observed
kinematics of NGC 6240. Obviously, this complicates attempts at
interpretation in terms of simple equilibrium models.

The main advantage of our data over the work of Tecza
\etal is that our spatial resolution of $\sim 0.2''$ is better than
their $0.8''$ ground-based seeing FWHM. However, our one-dimensional
coverage yields only limited insight as compared to the
two-dimensional stellar kinematical velocity field that Tecza
\etal obtained. The mean stellar velocity profile that we infer between 
the nuclei is not too dissimilar from the findings of Tecza et al.
However, they found in addition that there are considerable stellar
velocity gradients of $\sim 350 \kms$ peak-to-peak at the positions of
both of the nuclei. The kinematic major axis of these gradients do not
coincide with the line that connects the nuclei, and as a result,
these gradients are not evident in our data. Tecza \etal (2000)
interpreted their results as evidence for two rotating stellar disks
at the positions of each of the nuclei. While this is certainly a
possibility, an alternative view is that these gradients may represent
non-equilibrium dynamics associated with the galaxy merger (Mihos
2000).

\subsection{Kinematics of the Off-Nuclear Region}
\label{ss:kinoffnuc}

Bland-Hawthorn \etal (1991; hereafter BH91) determined the large scale
velocity field of the emission-line gas in NGC 6240. They found a
steep gradient in the gas velocity at a position $\sim 11.5''$ East
and $2.5''$ North of the nuclear region (white dot in
Figure~\ref{f:PCfield}). The velocity field there resembles a normal
disk rotation curve and they argued that this could be the center of
one of the two merging progenitor disks. With the assumption that the
gas is in circular rotation, they inferred the presence of $\sim
10^{11} \Msun$ of material at the position of the velocity
gradient. In the absence of a significant amount of starlight at this
position, they referred to this mass as a ``dark core''. They
discussed various possible physical interpretations of this result,
and suggested a super-massive black hole as the most plausible
one.\looseness=-2

We used the Royal Greenwich Observatory (RGO) Spectrograph with the
25cm camera on the AAT on May 9, 1991 to obtain new ground-based
long-slit emission-line spectra of NGC 6240. The R1200 grating was
used with a $1.3''$ slit, yielding a spectral resolution of $\sim 70
\kms$ FWHM.  The exposure time was 1800 seconds. The slit was centered
on the hypothesized black hole position and was placed along ${\rm PA}
= 155$ degrees (indicated by the white line in
Figure~\ref{f:PCfield}). This is along the direction of the steepest
velocity gradient reported in BH91. The spatial resolution of the AAT
observations was $\sim 1.5''$ FWHM. The inferred H$\alpha$+[NII]
velocity curve is shown in Figure~\ref{f:offnuccurves}a. The amplitude
of the AAT velocity gradient is similar to that seen in the
Fabry-Perot measurements of BH91 but is somewhat smaller than the
gradient seen in the Calar Alto observations of the latter
authors. This is probably due to differences in spatial resolution
among the observations. Either way, the new AAT spectra confirm the
existence of a steep velocity gradient that resembles a rotation
curve.

Identification of the most massive black holes in the nearby universe
is important for understanding both galaxy and black hole assembly
over cosmic time. The existence of bright quasars at $z\sim 6$
indicates that $\sim 10^9 \Msun$ black holes already existed early in
the Universe. Such black holes could have grown considerably through
subsequent accretion. However, developments since 1991 have made the
presence of a $\sim 10^{11} \Msun$ black hole at an off-nuclear
position in NGC 6240 seem unlikely. First, observations at different
wavelengths (optical, near-IR, X-rays) do not show any special feature
or source at the suggested black hole position. This is true also for
the HST broad-band and narrow-band imaging that we have presented
here.  Colbert, Wilson \& Bland-Hawthorn (1994) did not detect
significant radio emission (but super-massive black holes can of
course be radio-quiet). Second, it seems likely that the double
nucleus marks the centers of the two progenitor galaxies involved in
the merger, in particular now that it is known that both nuclei harbor
an AGN (Komossa \etal 2003). Third, $\sim 10^{11} \Msun$ seems
exceedingly massive for a black hole. The most massive black holes
known have masses of $\sim 3 \times 10^9
\Msun$ (Tremaine \etal 2002). Since black hole mass correlates well
with bulge mass and velocity dispersion, there is currently no
expectation that galaxies exist with central black holes as massive as
$\sim 10^{11} \Msun$.

Indepent of its origin, the off-nuclear velocity gradient remains
intriguing. We therefore decided to study it further using
$\Halpha$+[NII] emission-line spectra with STIS. The emission-line
flux in NGC 6240 decreases sharply with distance from the nuclear
region (see Figure~\ref{f:PCfield}) and it is quite low at the
position of the velocity gradient. To accumulate sufficient signal we
chose to use the $0.5''$-wide slit of STIS ({\tt
52x0.5}). Observations were obtained for seven parallel slit
positions, all placed along ${\rm PA} = 155$ degrees. One of the slits
was placed at the position reported by BH91. The other six slit
positions were obtained by moving the slit perpendicular to its length
(i.e., in the direction of position angle $245^{\circ}$) by distances
$y = -1.0''$, $-0.5''$, $+0.5''$, $+1.0''$, $+1.5''$ and $+2.0''$,
respectively. This yields contiguous coverage of a $3.5''$ wide area.

Several individual $380$ sec exposures were obtained for each slit
position. The number of exposures varied from 5 for $y = -1.0''$
(furthest from the nuclear region, where the emission line flux is
lowest) to 2 for $y = 2.0''$ (closest to the nuclear region where the
emission line flux is highest). To decrease the effects of read-out
noise, four-pixel on-chip binning was applied in both the spatial and
spectral directions (yielding effective pixels of $0.202''$ by
$2.24${\AA} $= 99.2 \kms$). The effective spatial apertures for the
observations are $0.2''$ (along to the slit) by $0.5''$ (perpendicular
to the slit). While the corresponding spatial resolution is admittedly
low for HST standards, it is still considerably better than the $\gta
1.5''$ resolution data that has been obtained from the ground. The
spectral resolution for an extended source observed with the
$0.5''$-wide slit is $\sim 105 \kms$ (Gaussian dispersion). This large
value does not cause problems because mean velocities of emission
lines can be determined with an accuracy that is smaller than
their width.

The analysis of the spectra proceeded similarly as described in
Section~\ref{ss:gasnuc}. It was found that the velocity dispersion of
the gas showed little systematic variation over the spatial region
under study, and was everywhere consistent with $\sigma \approx 250
\pm 50 \kms$. The mean velocity results shown in the figure were 
obtained by keeping $\sigma$ fixed to this value in the emission line
fits. Given the low $S/N$ of the data, it was found that this reduced
the noise in the inferred mean velocity profiles. The results for the
different slits were combined to form the two-dimensional map of the
mean gas velocity shown in Figure~\ref{f:offnucBH}a. The
``checkerboard'' appearance of the map is due to the fact that it is
built up from measurements for individual $0.2'' \times 0.5''$
``apertures''. The errors on the velocity measurements for the
individual apertures are in the range $50$--$100 \kms$. These
relatively large errors are due to the low $S/N$ of the spectra and
are responsible for the apparent ``graininess'' of
Figure~\ref{f:offnucBH}a. Nonetheless, the data provide useful new
insight into the velocity field.\looseness=-2

Figure~\ref{f:offnuccurves}b shows the projection of the data along
the $y$-axis of Figure~\ref{f:offnucBH}a, with additional 5-pixel
binning along the slit. In other words, it shows the kinematics for a
hypothetical $3.5''$-wide slit with $1.0''$ pixels, as function of
position along the slit. The projection and binning reduce the errors
in the mean velocity measurements, and the steep gradient in the gas
velocities is more clearly visible than in Figure~\ref{f:offnucBH}a.
Figure~\ref{f:offnuccurves}b shows the results for a much larger range
of $x$-values than Figure~\ref{f:offnucBH}a, to allow comparison with
the ground-based data. The STIS results are in good agreement with our
AAT data and with the Fabry-Perot data of BH91. The Calar Alto data of
the latter authors shows a somewhat larger velocity amplitude,
possibly due to a difference in the effective spatial
resolution. Either way, the datasets are all in adequate qualitative
agreement.

To provide further insight we compared the data to the expected
velocities for a model gas disk in rotation around black hole.  The
calculations were performed using the software described in van der
Marel \& van den Bosch (1998). It was assumed that the circular
velocity of the disk is dominated by the gravity of the black hole,
with negligible contributions from stellar gravity. This is motivated
by the fact that no concentration of stars is observed at the center
of the observed velocity gradient. The calculations take into account
the distance and inclination of the gas disk, the STIS PSF, and the
binning of the data into individual apertures. The models yield a
predicted line profile for each aperture, to which a Gaussian is
fitted for comparison to the actual data. For illustrative purposes,
we adopt an inclination of $i = 60^{\circ}$ for this model disk.

Models of rotation around a black hole always predict that the
observed velocity gradient increases as the spatial resolution is
improved. For the mass suggested by BH91, the models for the STIS
observational setup predict a major axis velocity difference of $1313
\kms$ between points at $\Delta x = \pm 0.2''$ from the 
black hole position. However, no gradient of such magnitude is seen in
our observed velocity field. The largest gradient in
Figure~\ref{f:offnucBH}a occurs at $(x,y) \approx (0.1'',0.5'')$ and
is only $\sim 490 \kms$ between points separated by $\Delta x =
0.4''$. Hence, we do not see the increase in velocity gradient with
improved spatial resolution that would be expected if a black hole
were present.

The black hole mass that produces roughly the same velocity gradient
as observed is only $6 \times 10^9 \Msun$. The velocity field
predicted by this model is shown in Figure~\ref{f:offnucBH}b. This
model provides a reasonable fit to the data along the major axis.
However, away from the major axis the fit is not very good. A generic
prediction of rotating gas disk models is that the velocity gradient
falls off steeply with distance from the major axis. This is clearly
seen in Figure~\ref{f:offnucBH}b. However, in the data the velocity
gradient is rather similar for all of the parallel slit positions
along which data were taken. This is illustrated further by
Figure~\ref{f:offnucBH}c,d, which are similar to
Figure~\ref{f:offnucBH}a,b, but have an additional Gaussian smoothing
with $0.5''$ dispersion applied to reduce the noise in the
data. Especially at values of $x > 0$, the isovelocity contours in the
data appear more nearly vertical than they do in the model
predictions.\looseness=-2

In summary, our HST observations do not support the hypothesis that
the off-nuclear velocity gradient in NGC 6240 is due to a black hole,
in agreement with other lines of evidence. Instead, it appears to be
more plausible to assume that the observed gradient is the combined
result of projection effects and non-equilibrium merger
kinematics. Mihos (2000) showed that gradients like the one observed
in NGC 6240 are observed in N-body simulations of mergers, due to
projection effects of multiple, distinct kinematic components. Also,
Figure~\ref{f:PCfield} shows that the position of the velocity
gradient lies across an ionized filament. So it is possible that the
kinematics are strongly influenced by a wind. Alternatively, the
kinematics could be affected by projection of the filament on top of
background or foreground emission, much like the situation of the
projected H$\alpha$ plume identified in the ULIRG IRAS19254-7245 (the
Superantennae; Mihos \& Bothun 1998).

\section{Conclusions}
\label{s:conc}

We have presented the results of an HST study of the ULIRG NGC 6240.
WFPC2 imaging was obtained in the broad B, V, and I-bands, as well as
in $\Halpha$+[NII] emission. A NICMOS K-band image from the HST Data
Archive was analyzed for comparison. STIS Ca triplet absorption-line
spectroscopy was obtained of the nuclear region, and $\Halpha$+[NII]
emission line spectroscopy was obtained of both the nuclear region and
an off-nuclear position where there is a steep velocity gradient.

The global morphology in the broad-band data shows a complex
morphology, consistent with the assumed merger nature of NGC 6240.
The emission-line gas shows a spectacular filamentary nebula. This
nebula appears consistent with a bipolar superwind, but other
interpretations can certainly not be ruled out on the basis of the
data presented here. Previous ground-based imaging had shown that NGC
6240 has a double nucleus. However, the nuclear morphology seen with
HST is considerably more complicated than what was inferred from
previous data. Optical images in both continuum light and emission
lines show three nuclear components, which we denote N1, N2, and
N3. The nucleus N1 corresponds to the northern nucleus seen in
ground-based images; the nuclei N2 and N3 jointly correspond to the
southern nucleus seen in ground-based images. The nuclei are spatially
resolved in the images. An unresolved component could have pinpointed
the site of an AGN, but such an unresolved component is not seen. The
southernmost nucleus N3 becomes less prominent in the HST data as one
progresses to redder wavelengths. This explains why Schulz \etal
(1993) found that the distance between the two nuclei seen in
ground-based data decreases with increasing wavelength. From our HST
data we conclude that N3 is not a true nucleus, but more likely an HII
region or an artifact of patchy dust obscuration. We estimate the
absorption $A_V$ towards the nuclei with a variety of simple
assumptions and obtain values that are broadly consistent with the
estimates obtained previously by other authors.  The distance between
N1 and N2 is larger than the distance between the radio and X-ray
nuclei of NGC 6240. In the K-band an additional component N1$'$
becomes evident close to N1. However, also the distance between N1$'$
and N2 is larger than the distance between the radio and X-ray nuclei
of NGC 6240. This implies that at least one of the radio/X-ray
components does not have an optical/near-IR counterpart, and
conversely, that at most one of the optical/near-IR counterpart
components has a radio/X-ray counterpart. In other words, even in
K-band we may not be seeing all the way through the dust.

The ionized gas velocity field in the nuclear region does not show
clear signs of rotation around either of the nuclei, or of equilibrium
motion in general. Both N1 and N2 appear to be located in regions of
lower mean velocity than their surroundings.  The velocity of N2 is
$\sim 250 \kms$ lower than that of N1. The highest mean velocity is
seen between N1 and N2, and at this point the mean velocity is $600
\kms$ larger than the mean velocity of N2. A similar velocity gradient 
between the nuclei was reported from observations of CO gas by Tacconi
\etal (1999). The velocity dispersion of the ionized gas is $\sim 300 \kms$ 
at both N1 and N2. The velocity dispersion between the nuclei is
lower and drops to values near $\sim 200 \kms$, larger than expected
for a cold disk. Also, the velocity dispersion map shows a complex
morphology. In gas kinematical studies of AGNs with HST one often sees
a very sharp increase in emission line width towards the AGN. Although
NGC 6240 harbors two AGNs (Komossa \etal 2003), no dramatic increase
in emission line width is seen towards either of the optical nuclei of
NGC 6240.

The stellar kinematics in the nuclear region are quite different from
the gas kinematics. The stars do not show the large velocity gradient
between the nuclei that is seen for both the ionized gas and for the
cold CO gas. Also, the mean stellar velocities of N1 and N2 are
similar, while the mean emission line velocity of N1 is $\sim 250
\kms$ lower than for N2. We cannot accurately measure the stellar
velocity dispersion in the nuclear region from the STIS absorption
line spectra. However, our low $S/N$ measurement $\sigma = 200 \pm 40
\kms$ is not inconsistent with the ground-based results of Tecza \etal
(2000). Both measurements indicate that the stellar velocity
dispersion in the nuclear region is lower than the values of $\sim 350
\kms$ that were obtained from earlier ground-based observations
(Lester \& Gaffney 1994; Doyon \etal 1994), probably due to
differences in spatial resolution.

We have also studied the emission-line velocity field at an
off-nuclear position where Bland-Hawthorn \etal (1991) reported a
steep velocity gradient. Both our ground-based AAT spectra and
HST/STIS spectra confirm the existence of the gradient. The STIS data
do no show a steepening of the gradient compared to the ground-based
data. Combined with detailed models and the lack of observed
counterpart at other wavelengths this suggests strongly that the
velocity gradient is not due to rotation around a black hole.

Overall, the spectroscopic data of NGC 6240 indicate that
line-of-sight projection effects, dust absorption, non-equilibrium
merger dynamics and the possible influence of a wind may be playing an
important role in the observed kinematics. In view of this, it is
unclear what the validity is of simple equilibrium models for the
observed kinematics (such as those advocated by, e.g., Tacconi \etal
1999 and Tecza \etal 2000), and whether the stellar velocity
dispersion can be used as an equilibrium measure of mass or
fundamental plane position (as done by Lester \& Gaffney 1994; Doyon
\etal 1994).


\acknowledgments
We would like to thank Zolt Levay for assistance with
Figure~\ref{f:PCfield} and David Zurek for assistance in the
preliminary reductions of the WFPC2 data. Support for proposals \#6430
and \#8261 was provided by NASA through a grant from the Space
Telescope Science Institute, which is operated by the Association of
Universities for Research in Astronomy, Inc., under NASA contract NAS
5-26555. JCM acknowledges support by the NSF through grant AST-9876143
and by a Research Corporation Cottrell Scholarship. This research has
made use of the NASA/IPAC Extragalactic Database (NED) which is
operated by the Jet Propulsion Laboratory, California Institute of
Technology, under contract with the National Aeronautics and Space
Administration.

\clearpage



\ifsubmode\else
\baselineskip=10pt
\fi


\clearpage


\ifsubmode\else
\baselineskip=14pt
\fi


\newcommand{\figcapDSS}{Image of NGC 6240 extracted from the Digitized Sky 
Survey. North is to the top and East is to the left. The cross marks
the position of the nuclear region. The square shows the size and
orientation of the WFPC2/PC chip field of view ($36'' \times
36''$). Figures~\ref{f:PCfield} and~\ref{f:PCblowup} are oriented such
that the black arrow points towards the bottom of the
page.\label{f:DSS}}

\newcommand{\figcapPCfield}{Composite WFPC2/PC image of NGC 6240. The 
field of view and orientation of the image are as indicated in
Figure~\ref{f:DSS}; North is at the top left corner of the image.  The
$\Halpha$+[NII] emission image is overlaid in yellow on a true-color
image of the continuum light. The latter was constructed from the
broad-band $B$, $V$ and $I$ images. The white dot markes the position
at which there is a steep gradient in the emission-line velocities
along a position angle of $155$ degrees (indicated by the white
line). The kinematics in this region are discussed in detail in
Section~\ref{ss:kinoffnuc}.\label{f:PCfield}}

\newcommand{\figcapPCblowup}{The nuclear region of NGC 6240. The image 
size for each panel is $4.55'' \times 4.55''$. The orientation is the
same as in Figure~\ref{f:PCfield} and is as indicated in
Figure~\ref{f:DSS}; North is at the top left corner of each
panel. {\bf (a)} True-color image constructed from the WFPC2
broad-band $B$, $V$ and $I$ images. {\bf (b)} WFPC2 $\Halpha$+[NII]
emission image. {\bf (c)} NICMOS $K$-band image. The smoother
appearance of the $K$-band image compared to the other two is due both
to the wider PSF at these wavelengths and the smaller importance of
dust absorption.\label{f:PCblowup}}

\newcommand{\figcapslitlayout}{Schematic representation of the 
slit layout for the STIS emission-line spectroscopy of the nuclear
region. Spectra were obtained for seven slit positions, as shown in
the figure and discussed in the text. The central slits have widths of
$0.1''$, the outer slits have widths of $0.2''$. The pixels along the
slits are indicated. Two-pixel on-chip binning was used for the
$0.2''$-wide slit observations while no on-chip binning was used for
the $0.1''$-wide slit observations. The nuclear components N1, N1$'$,
N2 and N3 of NGC 6240 are as indicated in
Figure~\ref{f:PCblowup}. North is up in this figure, by contrast to
Figures~\ref{f:PCfield} and~\ref{f:PCblowup}. The origin of the
coordinate system was arbitrarily chosen at N2.\label{f:slits}}

\newcommand{\figcapkinematics}{Two-dimensional grayscales maps of 
the $\Halpha$+[NII] emission line properties in the nuclear region of
NGC 6240, derived from HST/STIS spectroscopy: {\bf (a; top)} Total
flux; {\bf (b; middle)} mean velocity; {\bf (c; bottom)} velocity
dispersion. The Cartesian $(X,Y)$ coordinate system is centered on N2,
and has the $X$-axis along the line that connects N1 at
$(X,Y)=(-1.86'',0.0'')$ to N2. The flux distribution overall agrees
well with that seen in the $\Halpha$+[NII] WFPC2 narrow band image
(Figure~\ref{f:PCblowup}b, which is rotated with respect to the
present figure). The nuclei are clearly visible as peaks in the flux
distribution. The pixelized appearance of the flux map is due to the
fact that the nuclear region of NGC 6240 was mapped spectroscopically
with a set of disjunct ``apertures'', as shown in
Figure~\ref{f:slits}. By contrast, the mean velocity and velocity
dispersion maps were smoothed with a Gaussian of $0.1''$ dispersion,
to increase $S/N$. These maps have contours overlaid which are labeled
in km/s.\label{f:kinematics}}

\newcommand{\figcapgaskin}{$\Halpha$+[NII] emission line properties 
along a hypothetical $1.1''$-wide slit through N1 and N2, as function
of position along the slit. {\bf (a; top)} Total flux (in an area of
$0.1''$ along the slit by $1.1''$ perpendicular to the slit); {\bf (b;
middle)} mean velocity; {\bf (c; bottom)} velocity dispersion. Arrows
indicate the positions of the nuclei N1 and N2. These results were
obtained as described in the text, by co-adding the data for the
individual slit positions shown in Figure~\ref{f:slits}. This
essentially yields a projection of the contour plots in
Figure~\ref{f:kinematics} along the $Y$-axis of that figure.
\label{f:gaskin}}

\newcommand{\figcapstellarkin}{Results of Ca II triplet absorption 
line spectroscopy with a $0.2''$-wide slit along the position angle
that connects N1 and N2, as function of position along the slit. {\bf
(a; top)} Average continuum intensity over the observed wavelength
range (in an area of $0.1''$ along the slit by $0.2''$ perpendicular
to the slit). The positions of N1 and N2 are indicated. {\bf (b;
bottom)} Comparison of inferred mean stellar velocities (solid
symbols) to the mean emission line velocities copied from
Figure~\ref{f:gaskin} (open symbols). The stellar velocities differ
considerably from the gas velocities.\label{f:stellarkin}}

\newcommand{\figcapoffnuccurves}{Velocity curve 
of H$\alpha$+[NII] emission line gas at the off-nuclear position
indicated in Figure~\ref{f:PCfield}. Positive values of the abscissa
$x$ lie in the direction of ${\rm PA} = 155$ degrees.  {\bf (a; top
panel)} ground-based AAT data.  {\bf (b; bottom panel)} The HST/STIS
results of Figure~\ref{f:offnucBH} projected along the $y$ direction
with 5-pixel binning along the $x$ direction (corresponding to a
hypothetical $3.5''$-wide slit with $1.0''$ pixels). The results in
the two panels are in adequate agreement and confirm the steep
velocity gradient reported by Bland-Hawthorn \etal (1991; note that
their figure~2 has the horizontal axis flipped compared to the present
figure).\label{f:offnuccurves}}

\newcommand{\figcapoffnucBH}{Two-dimensional velocity field of emission line 
gas at the off-nuclear position indicated in Figure~\ref{f:PCfield}.
The $(x,y)$ coordinate system is centered on this position, and has
the positive $x$-axis in direction the direction of ${\rm PA} = 155$
degrees. Grey-scales run from $-225 \kms$ (black) to $+225 \kms$
(white), after subtraction of $7170 \kms$ (corresponding to the
symmetry point in Figure~\ref{f:offnuccurves}a). {\bf (a; top left
panel)} HST/STIS measurements for individual $0.2'' \times 0.5''$
``apertures''. {\bf (b; bottom left panel)} Model velocity field for a
disk at inclination of $60^{\circ}$ rotating around a $6 \times 10^9
\Msun$ black hole at $(x,y) = (0.1'',0.5'')$. {\bf (c; top right panel)} 
As panel (a), but smoothed with a Gaussian of dispersion $0.5''$. {\bf
(d; top right panel)} As panel (b), but smoothed with a Gaussian of
dispersion $0.5''$. The model that most resembles the data has a black
hole that is ten times less massive than was suggested by
Bland-Hawthorn \etal (1991), and the velocity field for this model
still differs considerably from the data, as discussed in the
text. Combined with other evidence, this suggests that the observed
velocity gradient is not due to a black hole but to the combined
result of projection effects and non-equilibrium merger
kinematics.\label{f:offnucBH}}


\ifsubmode
\figcaption{\figcapDSS}
\figcaption{\figcapPCfield}
\figcaption{\figcapPCblowup}
\figcaption{\figcapslitlayout}
\figcaption{\figcapkinematics}
\figcaption{\figcapgaskin}
\figcaption{\figcapstellarkin}
\figcaption{\figcapoffnuccurves}
\figcaption{\figcapoffnucBH}

\clearpage
\else\printfigtrue\fi

\ifprintfig


\clearpage
\begin{figure}
\epsfxsize=\hsize
\centerline{\epsfbox{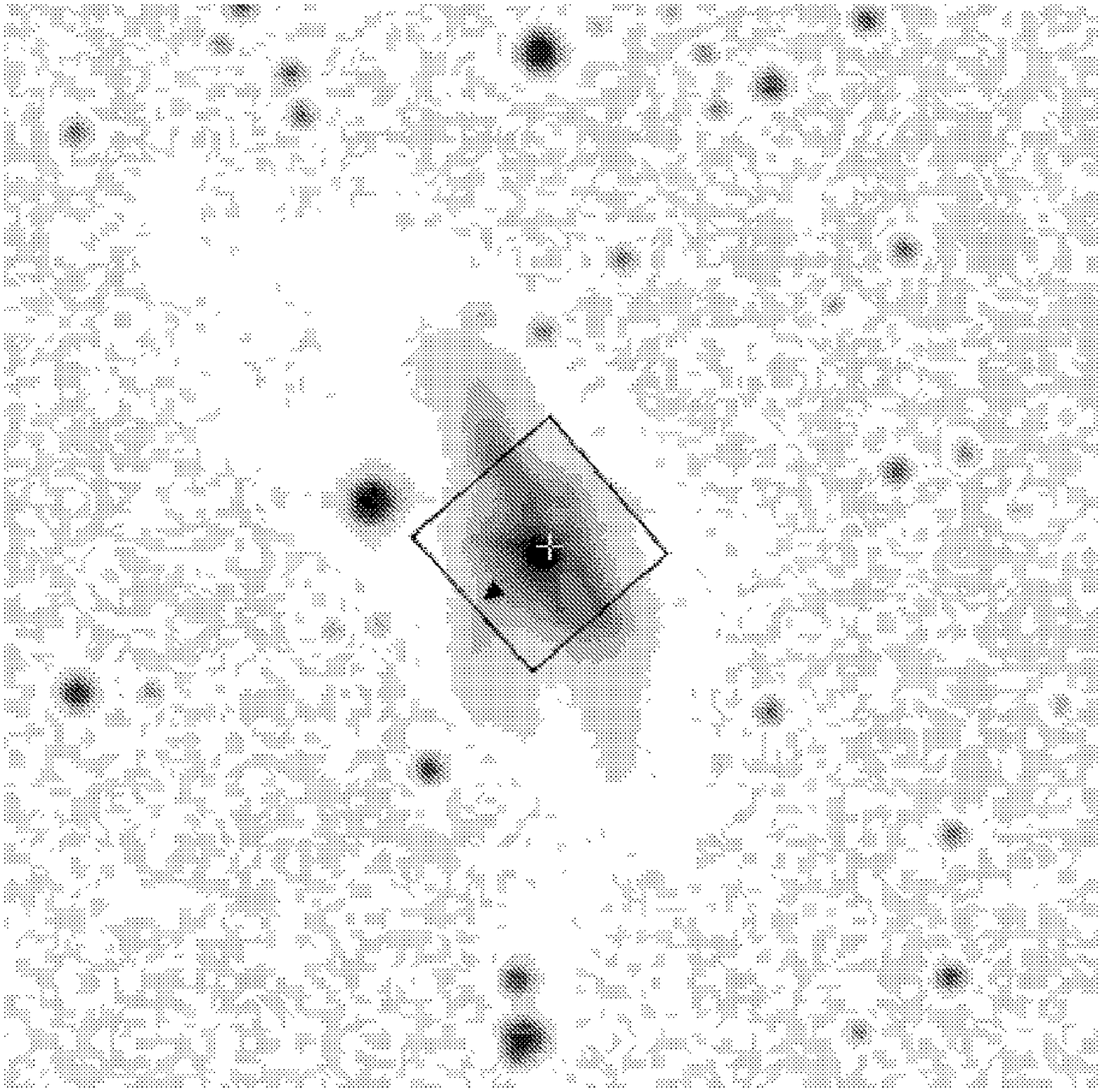}}
\ifsubmode
\vskip3.0truecm
\setcounter{figure}{0}
\addtocounter{figure}{1}
\centerline{Figure~\thefigure}
\else\figcaption{\figcapDSS}\fi
\end{figure}

\clearpage
\begin{figure}
\epsfxsize=\hsize
\centerline{\epsfbox{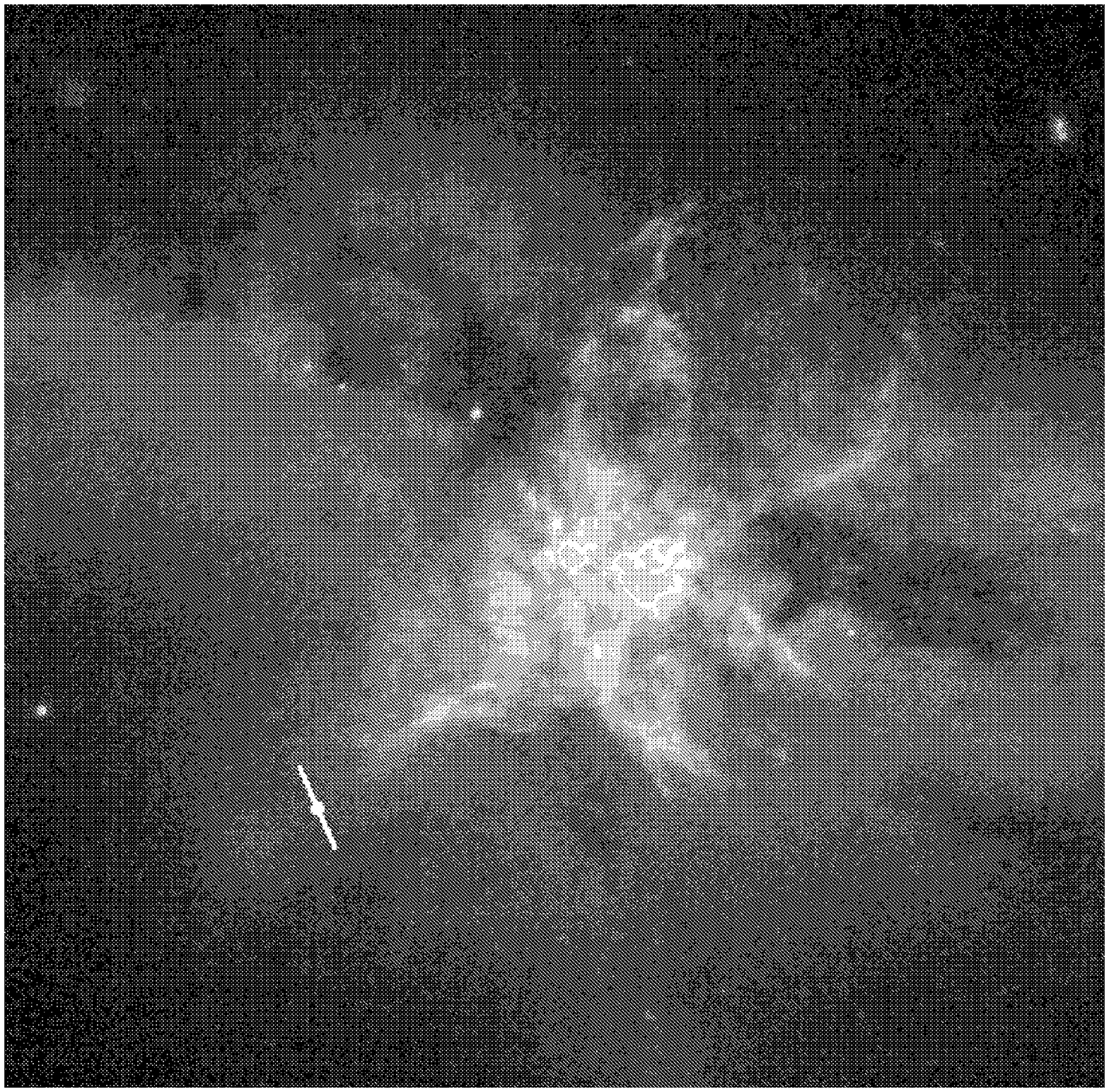}}
\ifsubmode
\vskip3.0truecm
\addtocounter{figure}{1}
\centerline{Figure~\thefigure}
\else\figcaption{\figcapPCfield}\fi
\end{figure}

\clearpage
\begin{figure}
\centerline{%
\epsfxsize=0.45\hsize
\epsfbox{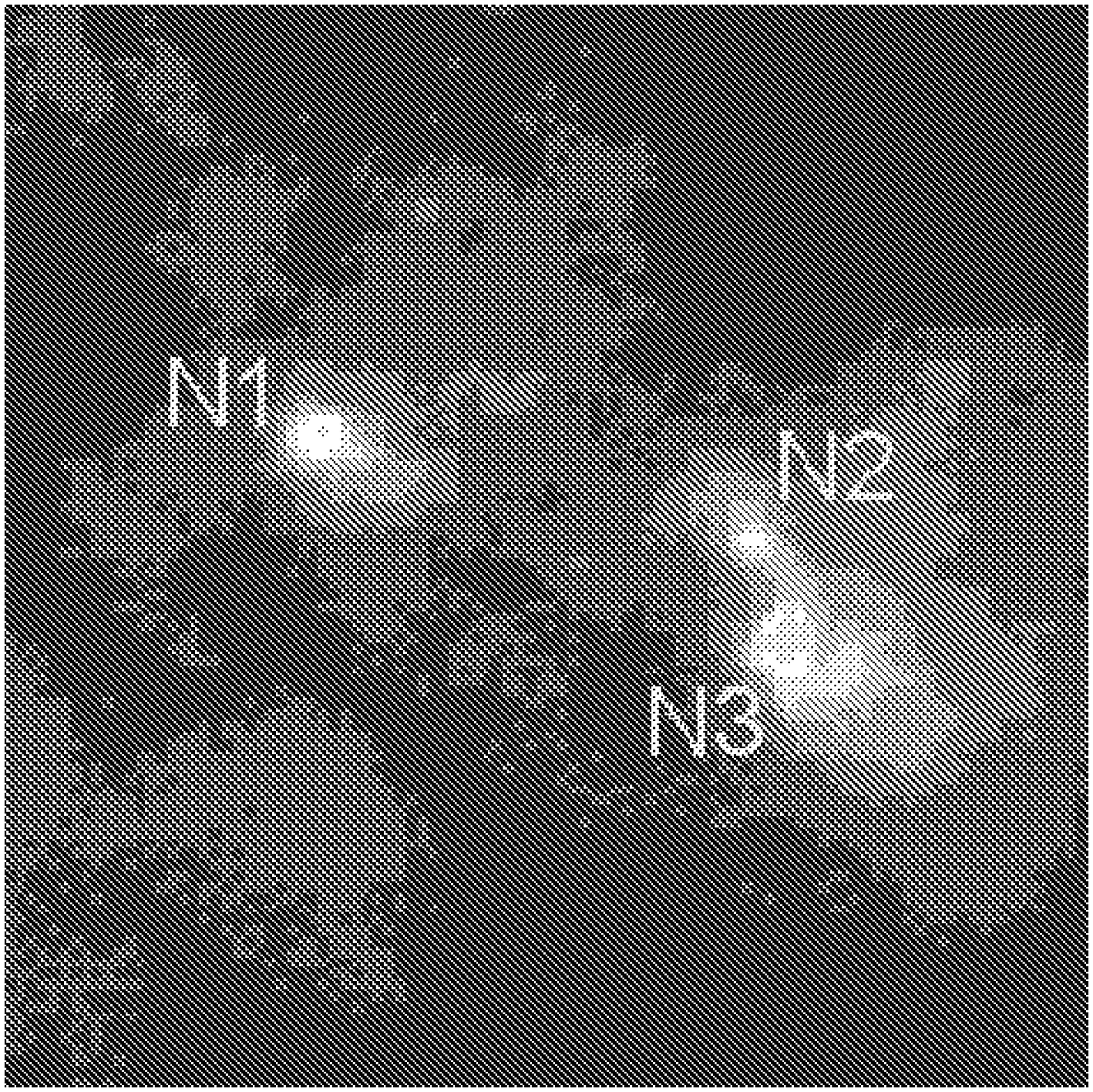}\quad
\epsfxsize=0.45\hsize
\epsfbox{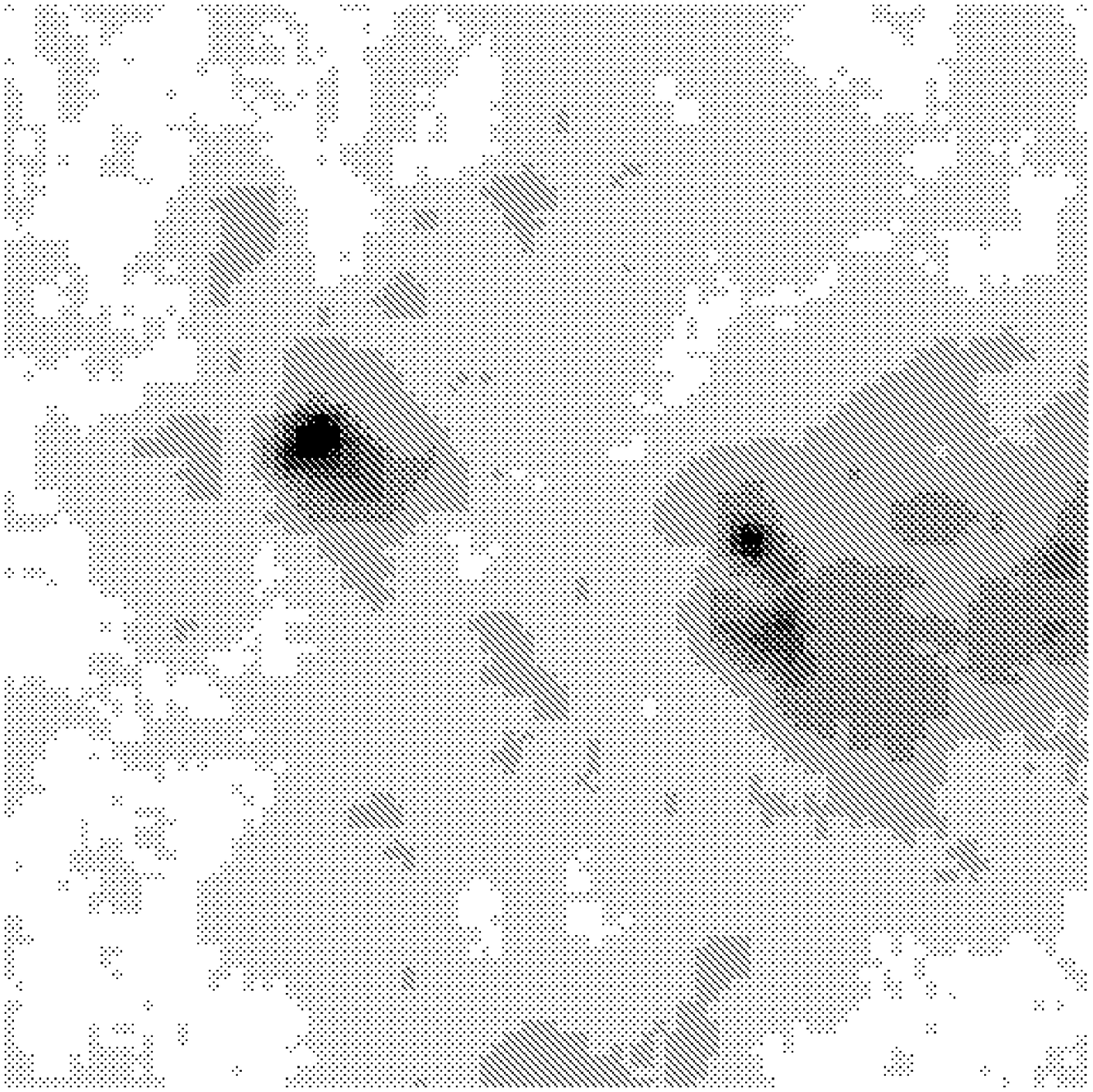}
}
\vspace{0.35truecm}
\centerline{%
\epsfxsize=0.45\hsize
\null\hspace{0.6truecm}\epsfbox{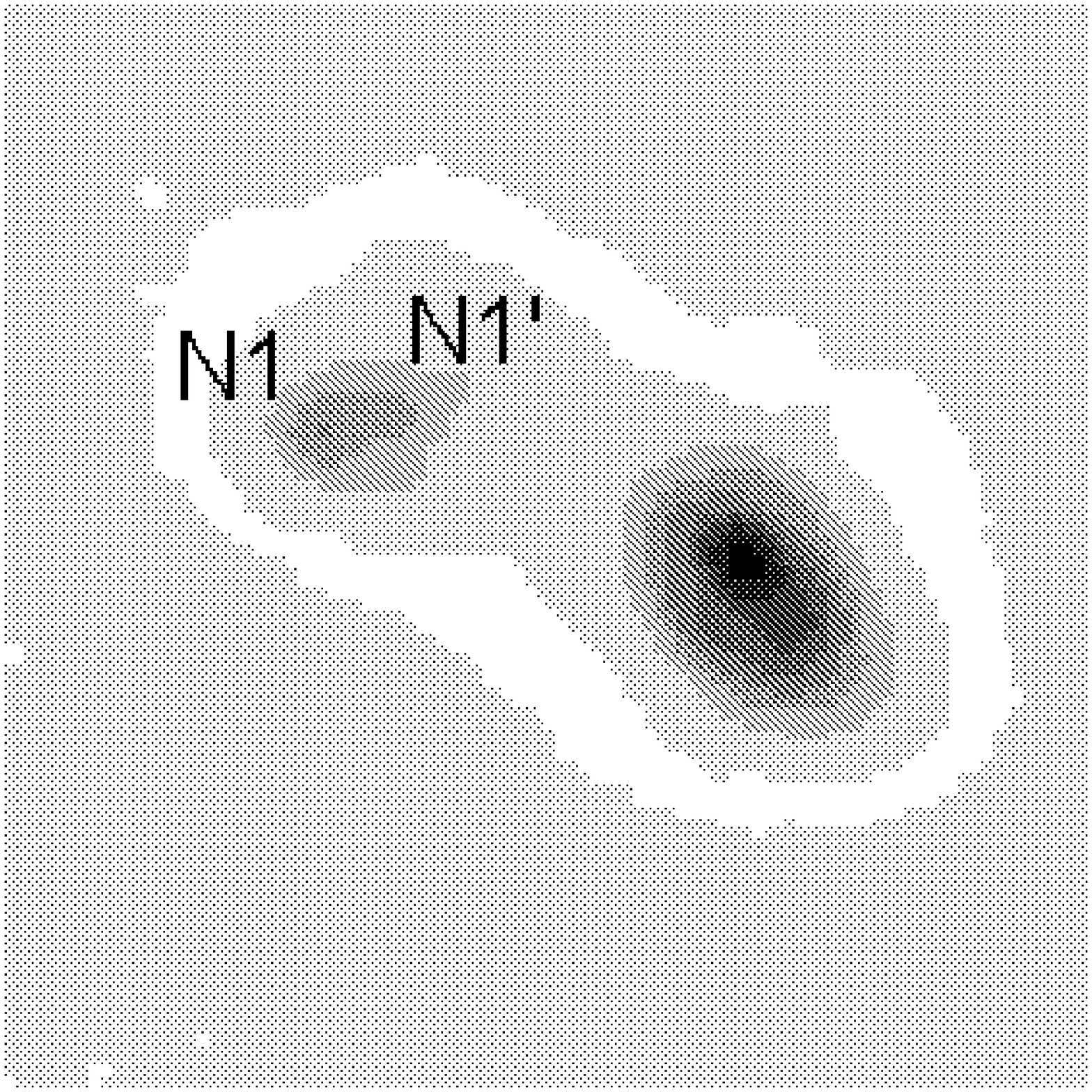}\hfill
}
\ifsubmode
\vskip3.0truecm
\addtocounter{figure}{1}
\centerline{Figure~\thefigure}
\else\figcaption{\figcapPCblowup}\fi
\end{figure}

\clearpage
\begin{figure}
\epsfxsize=0.8\hsize
\centerline{\epsfbox{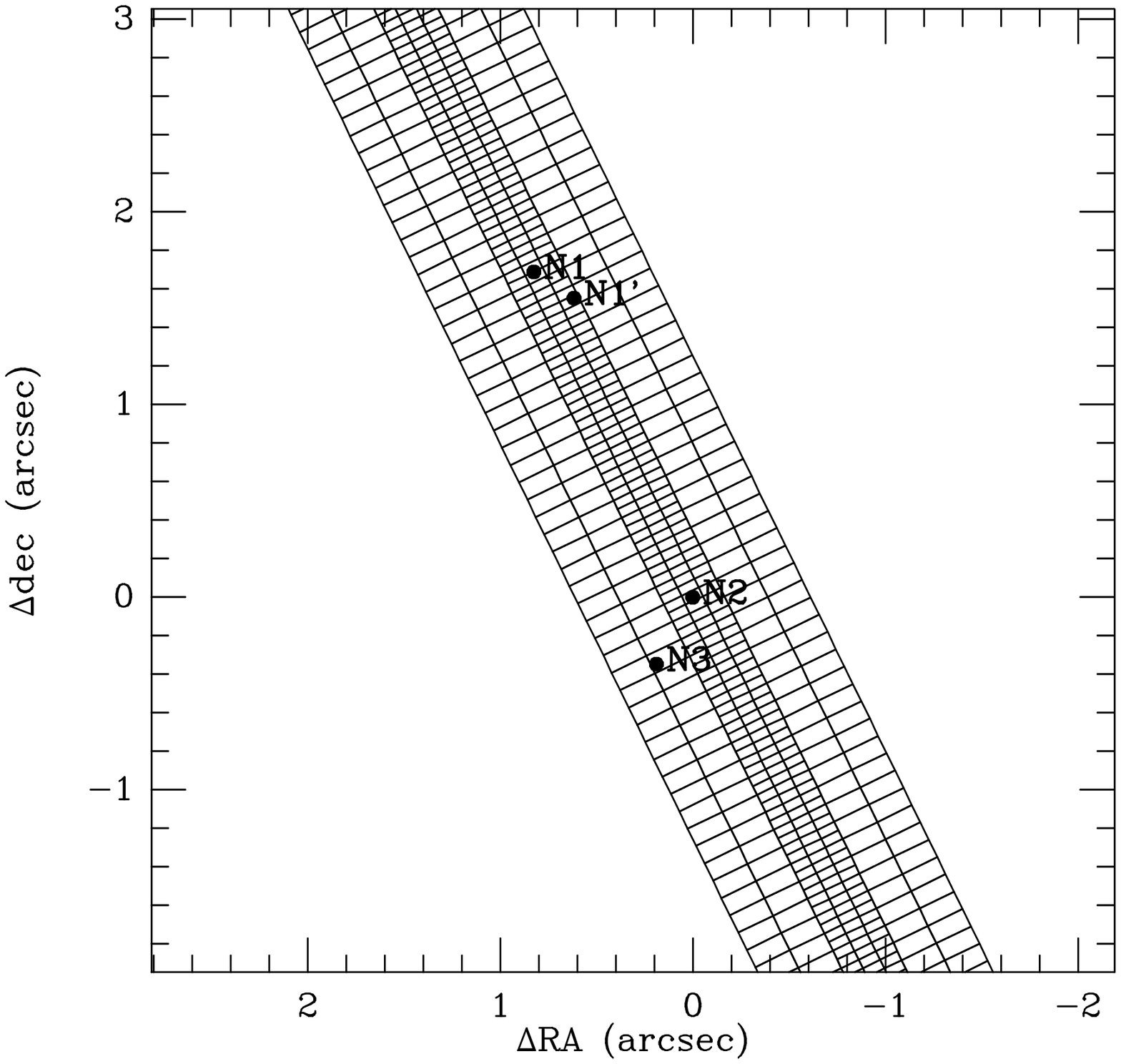}}
\ifsubmode
\vskip3.0truecm
\addtocounter{figure}{1}
\centerline{Figure~\thefigure}
\else\figcaption{\figcapslitlayout}\fi
\end{figure}

\clearpage
\begin{figure}
\epsfxsize=0.9\hsize
\centerline{\epsfbox{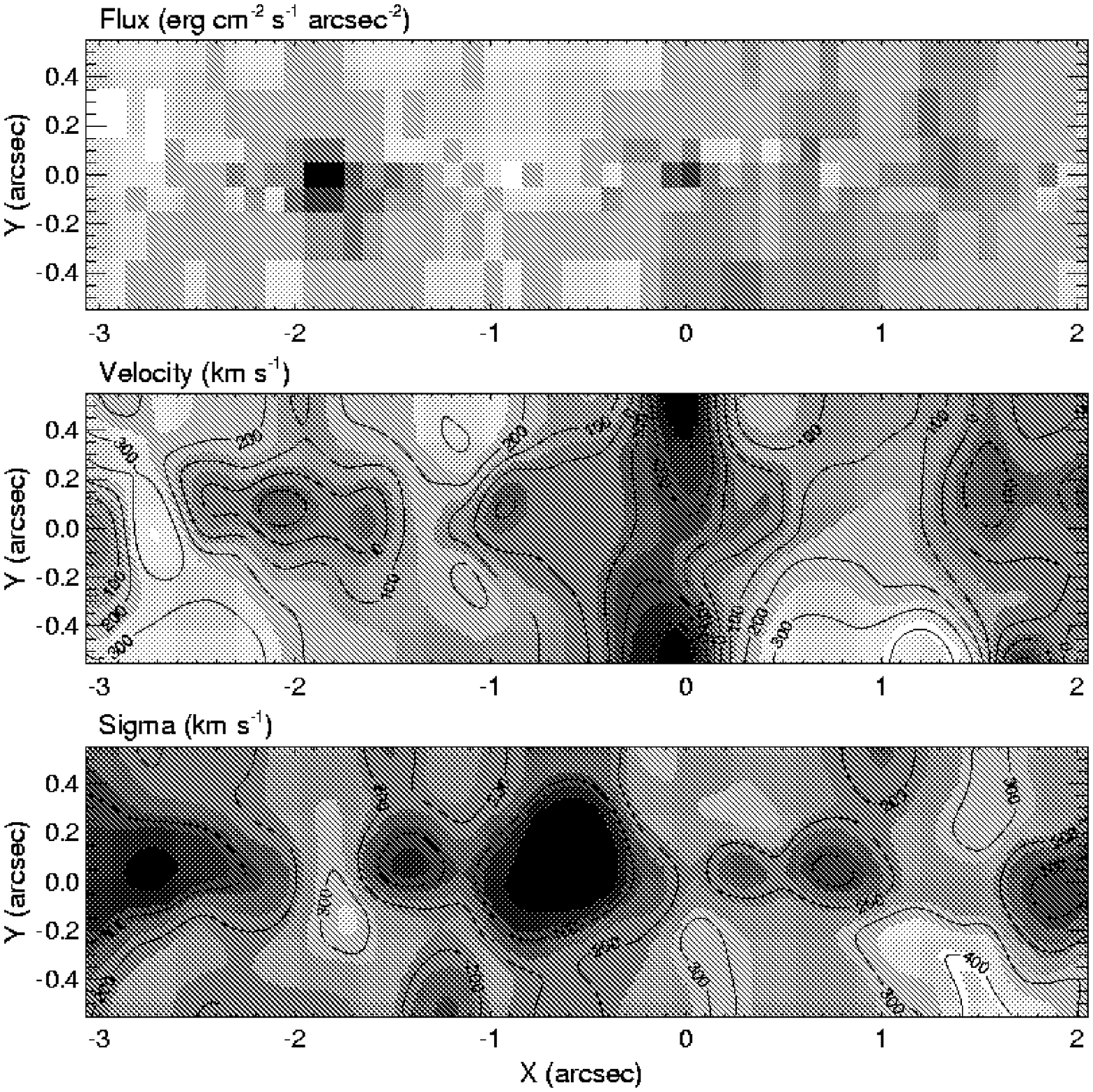}}
\ifsubmode
\vskip3.0truecm
\addtocounter{figure}{1}
\centerline{Figure~\thefigure}
\else\figcaption{\figcapkinematics}\fi
\end{figure}

\clearpage
\begin{figure}
\epsfxsize=\hsize
\centerline{\epsfbox{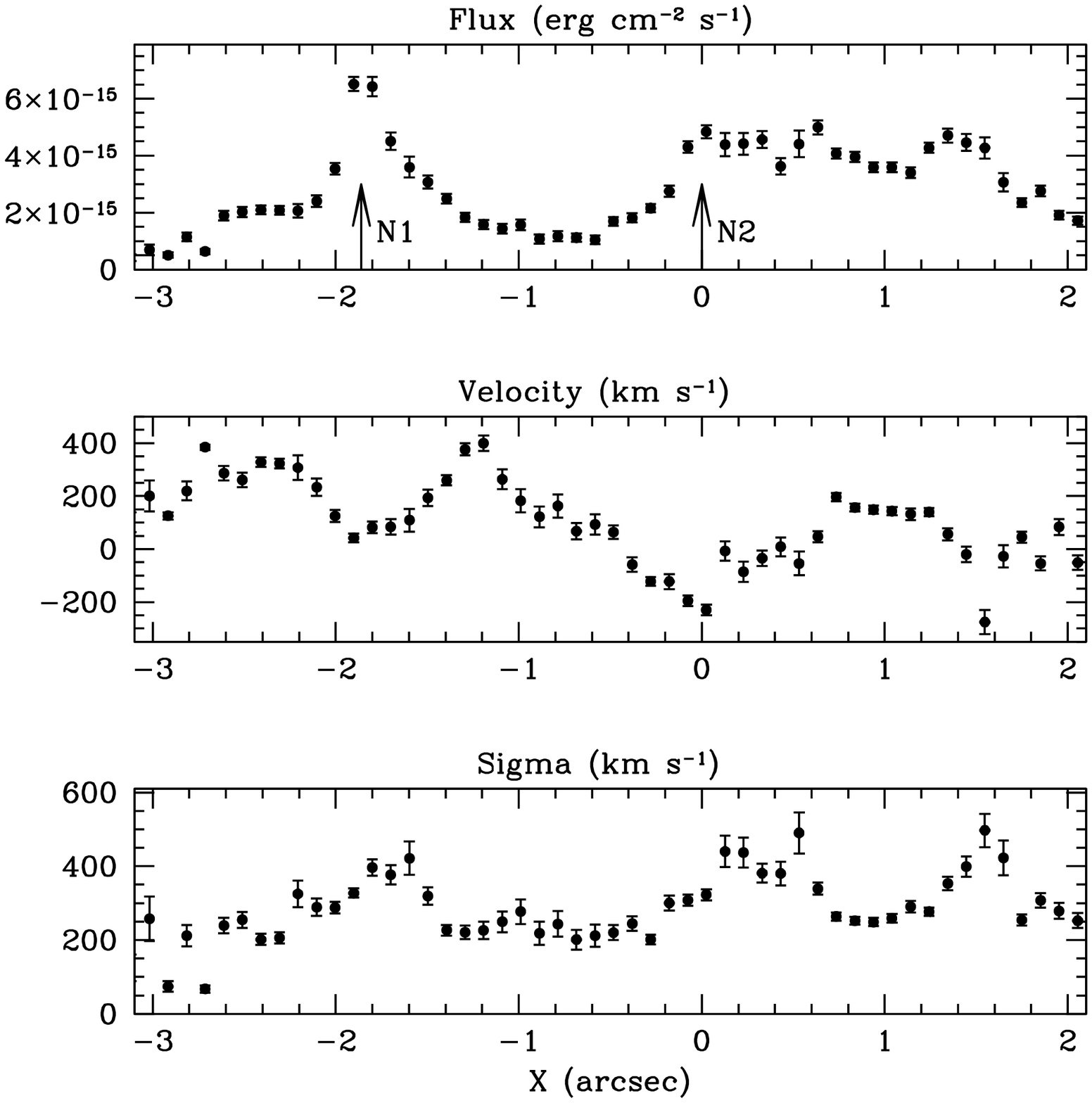}}
\ifsubmode
\vskip3.0truecm
\addtocounter{figure}{1}
\centerline{Figure~\thefigure}
\else\figcaption{\figcapgaskin}\fi
\end{figure}

\clearpage
\begin{figure}
\epsfxsize=\hsize
\centerline{\epsfbox{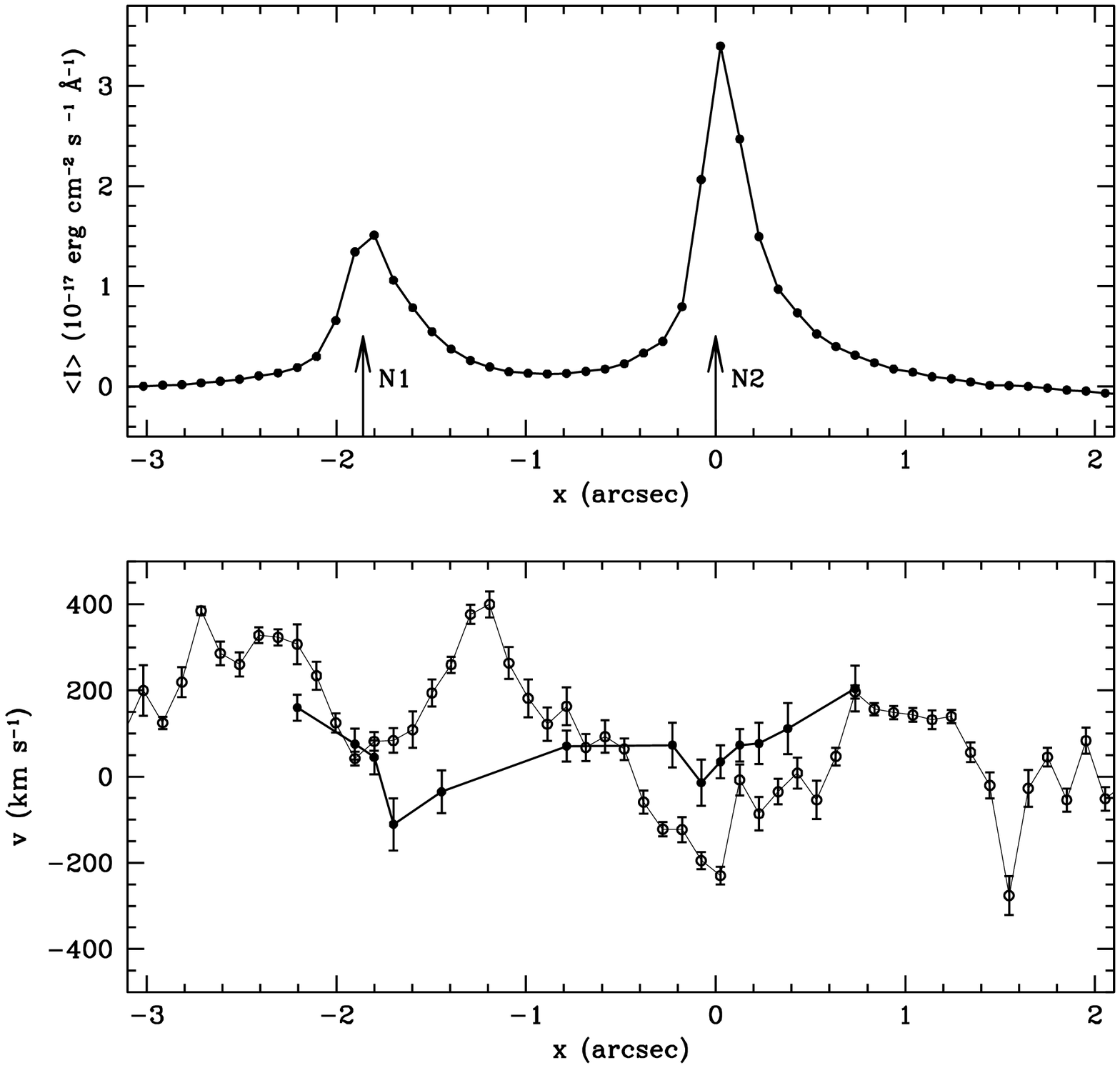}}
\ifsubmode
\vskip3.0truecm
\addtocounter{figure}{1}
\centerline{Figure~\thefigure}
\else\figcaption{\figcapstellarkin}\fi
\end{figure}

\clearpage
\begin{figure}
\epsfxsize=0.7\hsize
\centerline{\epsfbox{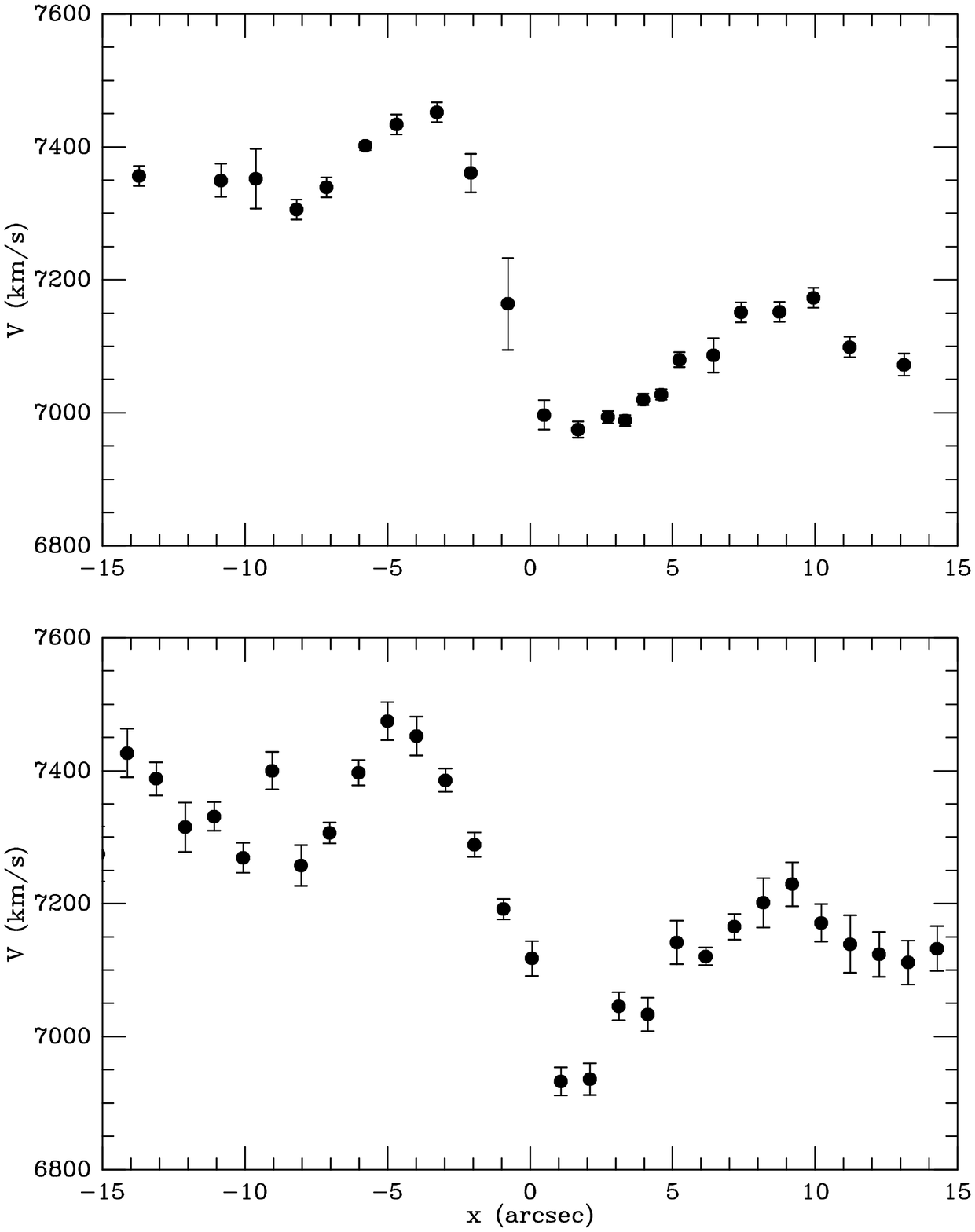}}
\ifsubmode
\vskip3.0truecm
\addtocounter{figure}{1}
\centerline{Figure~\thefigure}
\else\figcaption{\figcapoffnuccurves}\fi
\end{figure}

\clearpage
\begin{figure}
\epsfxsize=0.45\hsize
\hbox to \hsize{\noindent\epsfbox{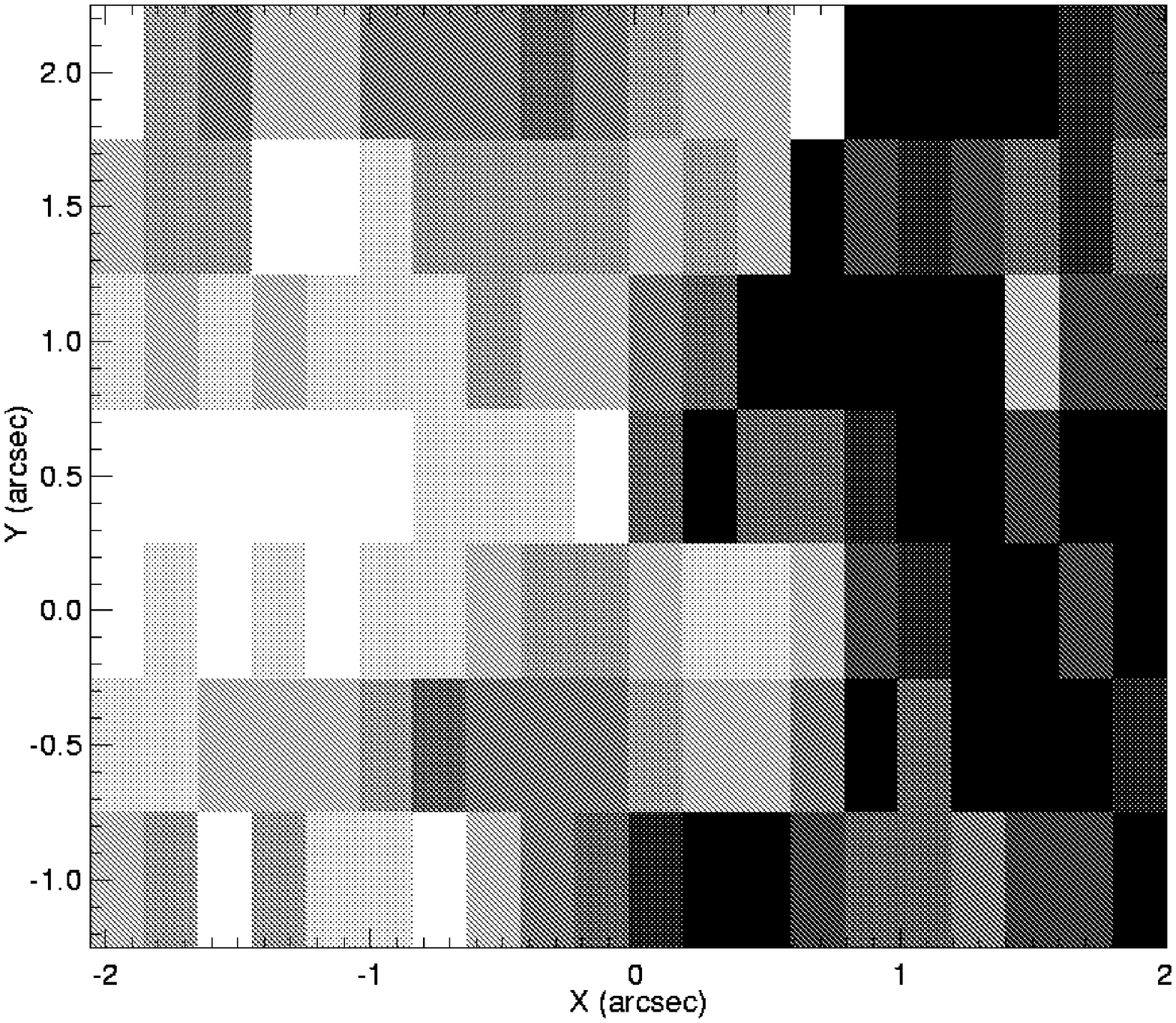}\hfill
\epsfxsize=0.45\hsize
\epsfbox{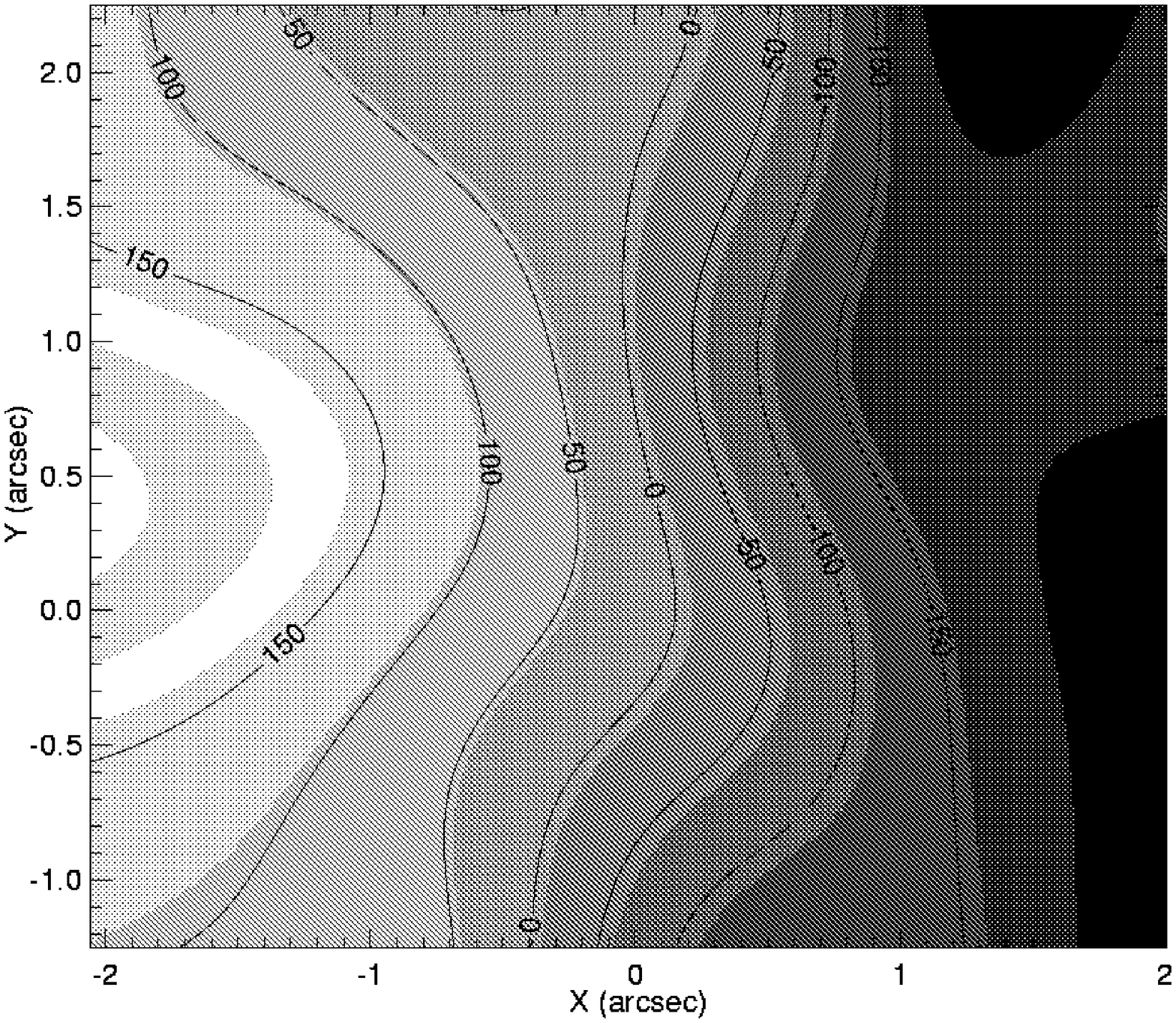}}
\smallskip
\epsfxsize=0.45\hsize
\hbox to \hsize{\noindent\epsfbox{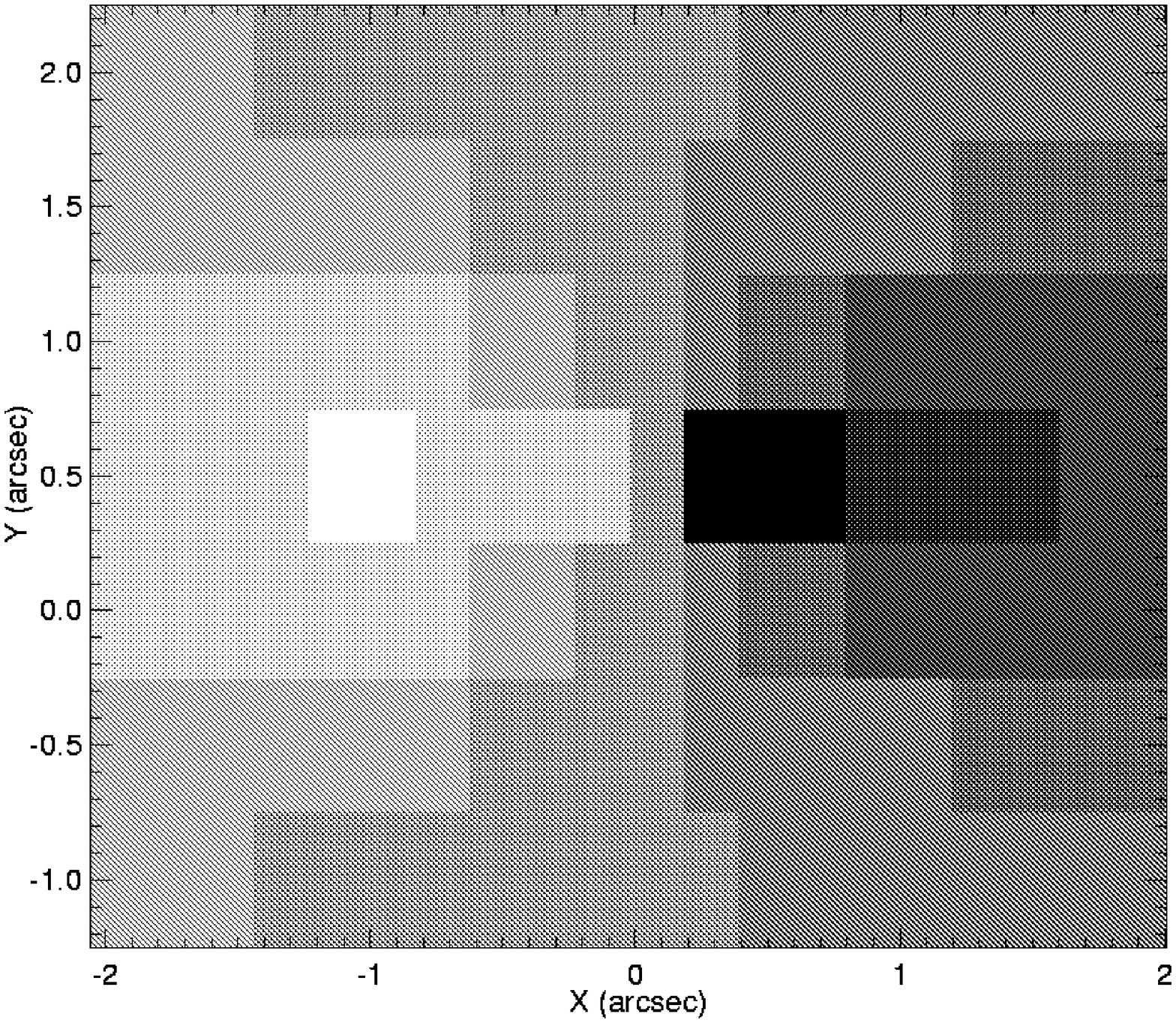}\hfill
\epsfxsize=0.45\hsize
\epsfbox{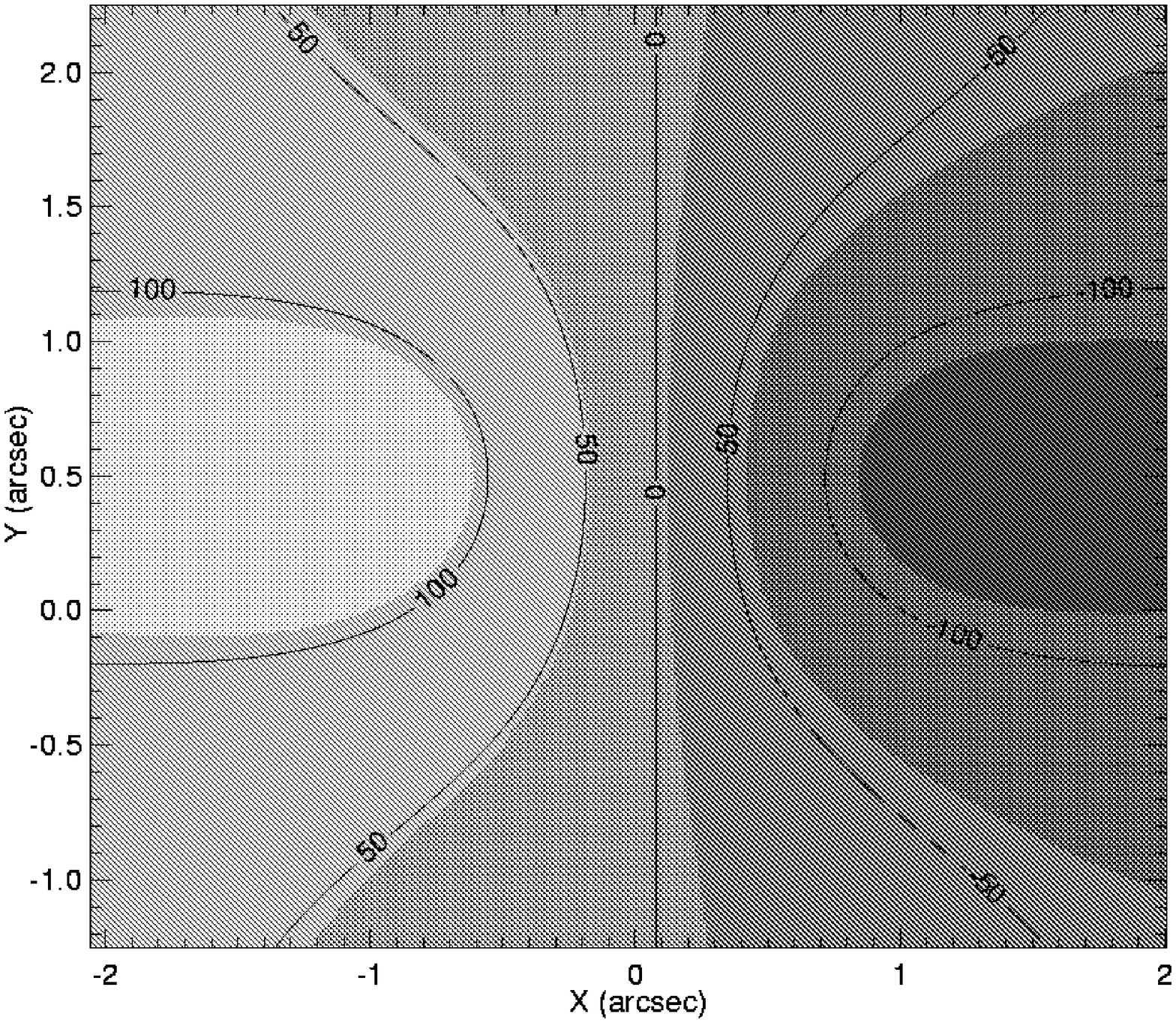}}
\ifsubmode
\vskip3.0truecm
\addtocounter{figure}{1}
\centerline{Figure~\thefigure}
\else\figcaption{\figcapoffnucBH}\fi
\end{figure}


\fi


\clearpage
\ifsubmode\pagestyle{empty}\fi


\begin{deluxetable}{crrrc}
\tablecaption{HST/WFPC2 images: observational setup\label{t:images}}
\tablehead{
\colhead{filter} & \colhead{$\lambda_0$} & \colhead{$\Delta \lambda$} &
\colhead{$T_{\rm exp}$} & \colhead{$N_{\rm exp}$} \\
 & \colhead{(\AA)} & \colhead{(\AA)} & \colhead{(sec)} & \\
\colhead{(1)} & \colhead{(2)} & \colhead{(3)} & 
\colhead{(4)} & \colhead{(5)}
}
\startdata
F450W & 4410 &  925 & 2100 & 3 \\
F547M & 5446 &  487 & 1200 & 3 \\
F658N & 6591 &   29 & 800  & 2 \\
F673N & 6732 &   47 & 2100 & 3 \\
F814W & 8203 & 1758 & 1200 & 3 \\
\enddata
\tablecomments{WFPC2 images of NGC 6240 were obtained with 5 different
filters. The filter name is listed in column~(1). Column~(2) and~(3)
list the central wavelength of the filter and the FWHM, as defined in
Biretta \etal (2001). Column~(4) lists the total exposure time per
filter, which was divided over the number of exposures listed in
Column~(5).}
\end{deluxetable}


\begin{deluxetable}{ccccccccc}
\tablecaption{Photometric properties of nuclear components\label{t:nucmags}}
\tablehead{
\colhead{nucleus} & \colhead{$B$} & \colhead{$V$} & \colhead{$I$} &
\colhead{$K$} & \colhead{$B-K$} & 
\colhead{$\langle A_V \rangle$} & \colhead{$V_{\rm corr}$} & \colhead{F(H$\alpha$+[NII])}\\
 & & & & & & & & \colhead{(erg cm$^{-2}$ s$^{-1}$)} \\
\colhead{(1)} & \colhead{(2)} & \colhead{(3)} & 
\colhead{(4)} & \colhead{(5)} & \colhead{(6)} &
\colhead{(7)} & \colhead{(8)} & \colhead{(9)} 
}
\startdata
N1 & 21.00 & 19.75 & 17.73 & 14.40 & 6.60 & $2.35^{+0.47}_{-0.33}$ & 17.40 & $6.9 \times 10^{-13}$ \\
N2 & 22.40 & 20.40 & 17.39 & 12.89 & 9.51 & $4.75^{+0.47}_{-0.33}$ & 15.65 & $3.7 \times 10^{-13}$ \\
N3 & 21.40 & 19.58 & 17.01 & 13.22 & 8.18 & $3.65^{+0.47}_{-0.33}$ & 15.93 & $3.7 \times 10^{-13}$ \\
\enddata
\tablecomments{Column~(1) is the identification of the nucleus as shown
in Figure~\ref{f:PCblowup}. Columns~(2)--(5) list the $B$, $V$, $I$ and $K$
magnitudes for a circular aperture with a radius of 4 pixels
($0.182''$). Column~(6) lists the implied $B-K$ color.
Column~(7) lists the average dust absorption $\langle A_V
\rangle $ with an error range, determined as described in the text.
Column~(8) lists the estimate $V_{\rm corr} = V - \langle A_V \rangle$
of the intrinsic $V$-band magnitude of each nucleus. Column (9) lists
the H$\alpha$+[NII] line flux of each nucleus.}
\end{deluxetable}



\end{document}